\documentclass[aps,prx,twocolumn,superscriptaddress]{revtex4-2}
\usepackage{rotating,graphicx}
\usepackage{gensymb}
\usepackage{physics}
\usepackage[dvipsnames]{xcolor}
\usepackage{appendix}

\begin{document}

\title{Virus-host interactions shape viral dispersal giving rise to distinct classes of travelling waves in spatial expansions}

\author{Michael Hunter}
\affiliation{Cavendish Laboratory, University of Cambridge, Cambridge, CB3 0HE, United Kingdom}

\author{Nikhil Krishnan}
\affiliation{Cavendish Laboratory, University of Cambridge, Cambridge, CB3 0HE, United Kingdom}

\author{Tongfei Liu}
\affiliation{Cavendish Laboratory, University of Cambridge, Cambridge, CB3 0HE, United Kingdom}
\affiliation{Department of Physics, University of Oxford, Oxford, OX1 3PJ, United Kingdom}

\author{Wolfram M\"{o}bius}
\affiliation{Living Systems Institute, University of Exeter, Exeter, EX4 4QD, UK}
\affiliation{Physics and Astronomy, College of Engineering, Mathematics and Physical Sciences, University of Exeter, Exeter, EX4 4QL, UK}

\author{Diana Fusco}
\email[]{E-mail: df390@cam.ac.uk}
\affiliation{Cavendish Laboratory, University of Cambridge, Cambridge, CB3 0HE, United Kingdom}

\date{\today}

\begin{abstract}
Reaction-diffusion waves have long been used to describe the growth and spread of populations undergoing a spatial range expansion. Such waves are generally classed as either pulled, where the dynamics are driven by the very tip of the front and stochastic fluctuations are high, or pushed, where cooperation in growth or dispersal results in a bulk-driven wave in which fluctuations are suppressed. These concepts have been well studied experimentally in populations where the cooperation leads to a density-dependent growth rate. By contrast, relatively little is known about experimental populations that exhibit density-dependent dispersal.

Using bacteriophage T7 as a test organism, we present novel experimental measurements that demonstrate that the diffusion of phage T7, in a lawn of host \textit{E. coli}, is hindered by steric interactions with host bacteria cells. The coupling between host density, phage dispersal and cell lysis caused by viral infection results in an effective density-dependent diffusion coefficient akin to cooperative behavior. Using a system of reaction-diffusion equations, we show that this effect can result in a transition from a pulled to pushed expansion. Moreover, we find that a second, independent density-dependent effect on phage dispersal spontaneously emerges as a result of the viral incubation period, during which phage is trapped inside the host unable to disperse. Additional stochastic agent-based simulations reveal that lysis time dramatically affects the rate of diversity loss in viral expansions. Taken together, our results indicate both that bacteriophage can be used as a controllable laboratory population to investigate the impact of density-dependent dispersal on evolution, and that the genetic diversity and adaptability of expanding viral populations could be much greater than is currently assumed.  
\end{abstract}

\maketitle

\section{Introduction}
Spatial range expansions are ubiquitous in nature, from the expansion of invasive plant species, through the migration of ancient human populations, to the range shifts of many organisms to higher altitudes and latitudes due to climate change \cite{Colautti2013, Cavalli-Sforza1993, Rosenberg2002, Ramachandran2005, Templeton2002, Chen2011, Hewitt1996, Hewitt2000}. One of the hallmarks of spatial expansions is the rapid loss of genetic diversity due to the enhanced fluctuations at the front~\cite{Hallatschek2008, Klopfstein2006}. This effect can, however, be significantly mitigated in the presence of density-dependent growth \cite{Saarloos2003, Roques2012}, such as an Allee effect \cite{Allee1931}, or density-dependant dispersal, where individuals in highly dense patches tend to disperse more quickly \cite{Birzu2019}. In particular, it has recently been shown theoretically that the ratio between the deterministic velocity of the front and that of its linearised approximation is sufficient to classify the expansions in three distinct types of travelling waves, nominally pulled, semi-pushed and fully-pushed, which respectively exhibit qualitatively distinct behaviors in the decay of heterozygosity, the stochastic wandering of the front position, and the probability distribution of the most recent common ancestor \cite{Birzu2018, Birzu2019}.

Because density-dependent growth can play such a crucial role in the evolutionary dynamic of a population, it has been extensively investigated in both naturally occurring range expansions in animals, such as the invasions of both Eurasian gypsy moths and house finches in North America \cite{Johnson2006, Veit1996}, and in laboratory microbial model systems, where the expansion dynamics in populations of the budding yeast \textit{Saccharomyces cerevisae} transition from pulled to pushed as growth becomes more cooperative \cite{Gandhi2016}, with a corresponding preservation of genetic diversity \cite{Gandhi2019}. In comparison, relatively little is known about the population dynamic of experimental systems that exhibit density-dependant dispersal, even if it has been documented in several natural populations~\cite{Matthysen2005} and the transition to pushed waves has been theoretically predicted~\cite{Birzu2019}.

One laboratory system that has been hypothesized to undergo density-dependent dispersal is bacteriophage expanding in a bacterial lawn. The crowded bacterial environment is thought to hinder phage diffusion because of steric interactions, resulting in a density-dependent diffusion coefficient due to the coupling between the host and the viral population densities \cite{Yin1992}. Direct experimental quantification of this density-dependent diffusion is, however, limited~\cite{Alvarez2007}, and its consequence on the front population dynamic mostly unknown.

Here, we address the open questions of (i) whether and how the rate of phage diffusion depends on the density of surrounding bacteria, (ii) under what conditions transitions to semi-pushed and fully-pushed expansions can occur, and (iii) what role density-dependent diffusion plays in this. We first design an experimental protocol to measure the effect of steric interactions between phage and the surrounding bacteria on phage dispersal. We then construct a system of reaction-diffusion equations to determine the phage front velocity, demonstrating that transitions to both semi-pushed and fully-pushed waves can occur. 
We find that the presence and location of these transitions are controlled by two independent effects that alter the density-dependent diffusion of the virus: the first is associated with the excluded-volume interactions with the surrounding bacteria, while the second spontaneously emerges from the viral infection dynamic, which prevents a viral particle from diffusing during infection of the host. Using stochastic agent-based simulations, we show that even the second effect alone, which applies to viruses beyond phage, can lead to a significant reduction in the rate of diversity loss in the viral population.

Taken together, our results identify bacteriophages as a controllable laboratory model system to investigate the role of density-dependent dispersal in evolution and provide a quantitative explanation of the physical mechanisms that control the phage population dynamic during a range expansion. Going beyond phages, our findings suggest that a broad range of viruses may expand via pushed travelling waves and, consequently, may be much more adaptable then previously thought.

\begin{figure*}[htb!]
\includegraphics[width=0.95\textwidth]{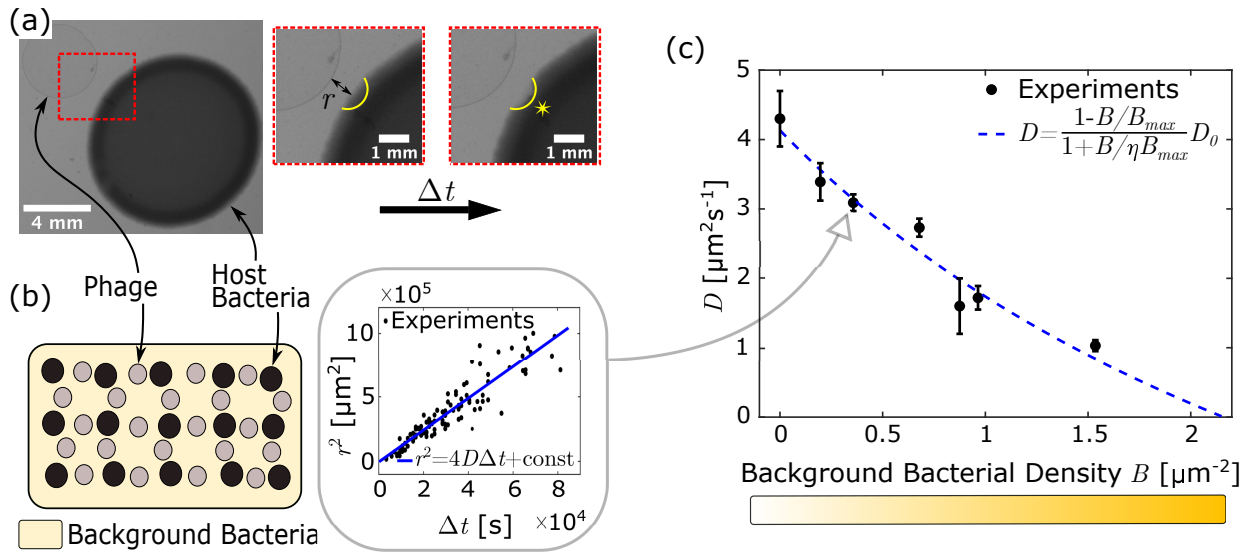} 
\caption{\label{fig:Diffusion_Experiment} Experiments show how rate of phage diffusion is reduced by surrounding bacteria. (a): The basis of the experimental set-up, consisting of a droplet of phage and host bacteria, separated by a distance $r$. After time $\Delta t$, a plaque begins to form in the host bacterial droplet (starred region). (b): The full experimental set-up, consisting of many phage-host droplet pairs on top of a lawn of phage resistant bacteria of variable density. The presence of chloramphenicol in the plate media ensures that the background bacterial density is constant over the course of the experiment (see Methods). An example plot of $r^2$ against $\Delta t$ data, with linear fit, for a resistant bacteria density of 0.36 $\mu$m$^{-2}$ is shown. (c): The diffusion coefficient obtained as a function of resistant bacteria density (i.e. for several instances shown in (b)), fit with Fricke's Law (Eq.~\ref{eqn:hindered_diffusion}).}
\end{figure*}

\section{Experimental Measurements of Density-Dependent Dispersal in Coliphage T7}
Starting with Yin and McCaskill \cite{Yin1992}, it is usually recognized that bacteria can act as a barrier to phage diffusion, resulting in a diffusion coefficient that depends on the bacterial density. This dependence is indeed necessary in phage expansion models to correctly reproduce the non-monotonicity of front velocity observed as a function of bacterial density \cite{Yin1992, You1999, Fort2002, Ortega-Cejas2004}. While it has been shown that phage diffuses faster in the bulk of a plaque than at the edge~\cite{Alvarez2007}, the dependence of phage diffusion coefficient on bacterial density has never been quantitatively measured. 

To address this need, we designed an experimental setup where (i) host density can be quantified and maintained uniform in space and constant in time, and (ii) the effect of steric interactions is decoupled from the viral infection dynamic. To this end, we moved away from classic plaque-in-agar assay, which exhibit a fragmented host distribution at the microscopic level, and built a uniform bacterial lawn by directly pouring an \textit{E. coli} liquid culture of known density on top of 2\% agar LB plates containing chloramphenicol (background bacteria in Fig.~\ref{fig:Diffusion_Experiment}b, Methods). These bacteria are susceptible to the antibiotic, which prevents their growth ensuring a constant host density during the experiment, and are engineered to prevent phage adsorption~\cite{Mobius2015, Qimron2006}, so as to serve as passive barrier to phage dispersal. Phage droplets were then inoculated across the lawn (grey in Fig.~\ref{fig:Diffusion_Experiment}b) at different distances from droplets of a second \textit{E. coli} strain, susceptible to phage and resistant to chloramphenicol (black in Fig.~\ref{fig:Diffusion_Experiment}b). The time $\Delta t$ required by the phage to travel the distance $r$ between a viral droplet and a close-by susceptible bacteria droplet was monitored \textit{in vivo} by tracking the appearance of clearings in the susceptible droplets (Fig.~\ref{fig:Diffusion_Experiment}a and b).

By gathering statistics over many droplet-droplet pairs, we were able to first confirm that the relationship between distance travelled and mean first passage time is consistent with diffusive behavior for the whole range of background densities tested (Fig.~\ref{fig:Diffusion_Experiment}b), and then calculate the rate of phage diffusion $D$ as a function of background bacterial density (Fig.~\ref{fig:Diffusion_Experiment}c). Additional tests were also performed to confirm that phage did not significantly diffuse out of the plane and into the agar during the course of the experiment (Appendix~\ref{app:agar_diffusion}).

Building on previous efforts to account for density-dependence in plaque models \cite{Yin1992}, we fit our data using Fricke's Law \cite{Fort2002, Fricke1924}, which describes the diffusion of a solute through a suspension of spheroids \cite{Crank1979}:
\begin{equation}\label{eqn:hindered_diffusion}
    D = \frac{1-b}{1+\frac{b}{\eta}}D_0 \quad\mathrm{;}\quad b=\frac{B}{B_{max}},
\end{equation}
where $b$ indicates the fraction of bacteria $B$ relative to a maximum value $B_{max}$ and $\eta$  accounts for the shape of the cells: spherical cells correspond to $\eta$=2, while \textit{E. coli} cells have previously been determined to correspond to $\eta$=1.67 \cite{Fort2002}. Our experimental data allow for the first time to estimate the two fitting parameters required by Frickes's law in this context: the free diffusion coefficient $D_0$ (i.e. the diffusion coefficient in the absence of surrounding bacteria), and the bacterial density $B_{max}$ at which diffusion is expected to be completely halted. We estimate $D_0=4.13\pm0.19\mu \textrm{m}^2\textrm{/s}$, which is in good agreement with the rate of 4 $\mu \textrm{m}^2 \textrm{/s}$ previously determined by Ouchterlony double immunodiffusion in 10 g/l agar of phage P22 (similar size and shape of T7) \cite{Stollar1963, Ackermann1976}; and $B_{max}=2.16\pm0.19\;\mu \textrm{m}^{-2}$, which is consistent with the typical dimensions of an \textit{E. coli} cell (assuming \textit{E. coli} cells are approximately 0.5 $\times$ 2 $\mu$m, we would expect a 1 $\mu$m$^2$ cross section to contain between 1 and 4 closely packed cells, depending on their orientation and deformation). Note that while we will use the expression in Eq.~\ref{eqn:hindered_diffusion} to account for the effect of steric interactions in phage dispersal on a uniform bacterial lawn, this relationship can be applied to any scenario in which the bacterial density distribution is known, even if it is non-uniform.

\section{Modelling Plaque Growth: Density-Dependent Diffusion and Adsorption to Infected Cells}
To investigate whether the phage expansion on a bacterial lawn occurs as a pulled or a pushed wave, and to uncover the role of host density-dependence, we compare the actual front velocity with the velocity $c_F$ of the corresponding linearised system, as their ratio has been shown to be sufficient to determine the wave class in single species range expansions~\cite{Birzu2018}. To this end, we develop a mathematical model that accommodates the density-dependent diffusion we have experimentally measured.

We model the spatial dynamics of bacteriophage plaque growth by considering the interactions between three populations: viruses (phage) $V$, uninfected host bacteria $B$ and infected host bacteria $I$, similar to \cite{Yin1992, Fort2002, Jones2012, You1999, Ortega-Cejas2004, Amor2010, Amor2014, Rioja2015}. The process may be summarised as 
\begin{equation}
    V+B \xrightarrow[\alpha]{\textrm{rate}} I \xrightarrow[\tau]{\textrm{delay}} \beta V,
\end{equation}
where $\beta$ is the burst size, $\alpha$ is the rate of adsorption, and $\tau$ is the lysis time.

As the model is deterministic, without loss of generality, we describe these populations with a set of reaction-diffusion equations in 1D, similar to those examined by Jones \textit{et al.} \cite{Jones2012}:
\begin{subequations}\label{eqn:Our_Model}
\begin{equation}
\pdv{B}{t} = -\alpha VB,
\end{equation}
\begin{equation}
\pdv{I}{t} = \alpha VB - \alpha V_{t-\tau}B_{t-\tau},
\end{equation}
\begin{equation} 
\pdv{V}{t} = \pdv{}{x} \left(D \pdv{V}{x} \right) - \alpha VB - \alpha^* VI + \beta\alpha V_{t-\tau}B_{t-\tau},
\end{equation}
\end{subequations}
where $V$, $B$ and $I$ indicate the concentration of the population as a function of space and time. The subscript is used to indicate that those terms are evaluated at time $t-\tau$. $D$ is the density-dependent diffusion coefficient of the phage, determined from fitting Fricke's law to experimental results in the previous section (Eq.~\ref{eqn:hindered_diffusion}). $\alpha^*=\alpha$ or $\alpha^*=0$, depending on whether adsorption to previously infected hosts is allowed or prevented, respectively. We assume that the host bacteria are motionless and that adsorption to uninfected hosts always leads to successful infection while neglecting desorption.

\begin{figure*}
\includegraphics[width=0.99\textwidth]{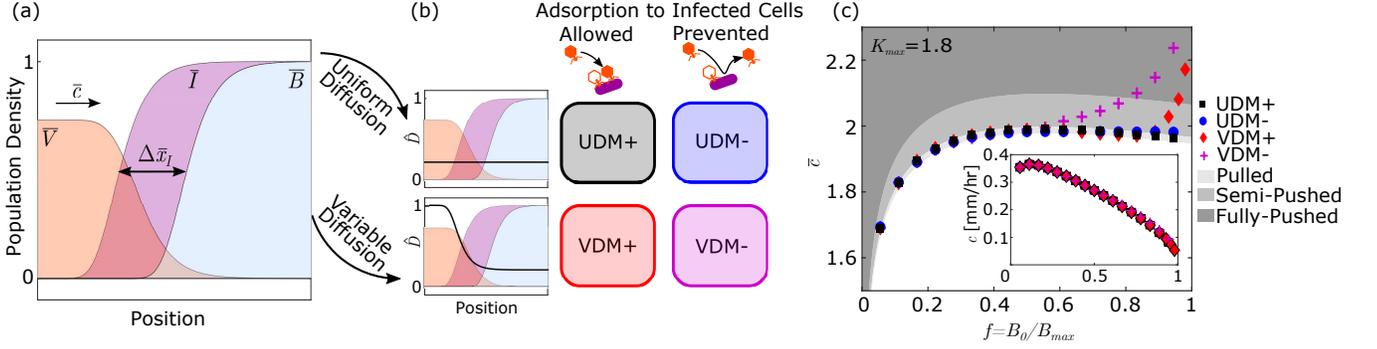}
\caption{\label{fig:fronts_transitions} (a): A sketch of the population concentrations $\overline{B}$, $\overline{I}$ and $\overline{V}$ as a function of location at the expansion front (the precise location is not important here, as the qualitative shape of the fronts remain constant during the expansion). The front is propagating with dimensionless velocity $\overline{c}$ to the right. The dimensionless width $\Delta\overline{x}_I$, characterising the width of the infected region is given by the difference in position of the uninfected ($\overline{B}$) and infected fronts ($\overline{B}+\overline{I}$). (b): The differing diffusion and adsorption behaviours explored lead to four different model variants in this work. Variants either have a Uniform or Variable diffusion coefficient $\hat{D}=D/D_0$ (UDM or VDM respectively, black line in (b)), and adsorption to previously infected cells either does (+), or does not (-) occur, leading to the four model variants (UDM+, UDM-, VDM+ and VDM-). (c): Dimensionless front velocity $\overline{c}$ as a function of bacteria fraction $f$, with shaded regions indicating the different expansion types. Error bars on the velocities are smaller than the symbols. Inset also shows the dimensional velocity $c$. Model parameters are chosen to represent typical T7 expansions with $\beta=50$, $\tau$=18 mins, and $\alpha B_{max}$=0.1 min$^{-1}$~\cite{Yin1992, Jones2012, Rioja2015}, corresponding to $K_{max}=1.8$ in our model.}
\end{figure*}

Our model introduces two ingredients that are biologically and physically relevant, and that are expected to affect the front dynamic. First, in contrast to previous work \cite{Yin1992, Fort2002, You1999, Ortega-Cejas2004, Amor2010, Amor2014, Rioja2015}, where the diffusion coefficient $D$ only depends on the initial bacterial density $B_0$ ($b=\frac{B_0}{B_{max}}$ in Eq.~\ref{eqn:hindered_diffusion}), we allow $D$ to vary in time and space according to the local bacterial density ($b=\frac{B+I}{B_{max}}$ in Eq.~\ref{eqn:hindered_diffusion}), resulting in faster diffusion inside the phage clearing (Fig.~\ref{fig:fronts_transitions}a,b). Secondly, we allow for the possibility that phage can adsorb to previously infected cells ($-\alpha^* VI$ term in Eq.~\ref{eqn:Our_Model}c), as is the case for phage T7. The presence or absence of these two effects generates four model variants that are summarized in Fig.~\ref{fig:fronts_transitions}b: Uniform vs. Variable Diffusion model (UDM vs. VDM), and adsorption vs. non-adsorption to infected cells (+ vs. -).

In line with previous studies, we cast the equations using dimensionless variables.
We measure concentrations in terms of the initial bacterial density $B_0$, time in units of $\tau$, and length in units of $L=\sqrt{D(B_0)\tau}$ (spatial scale of diffusion at the front within the lysis time). This results in the following set of dimensionless variables: $\overline{B}\equiv B/B_0$, $\overline{I}\equiv I/B_0$, $\overline{V}\equiv V/((\beta-1)B_0)$, $\overline{t}\equiv t/\tau$, $\overline{x}\equiv x/L$ and $K\equiv \alpha \tau B_0$. Consequently, $\overline{c}=c\sqrt{\tau/D}$, where $\overline{c}$ and $c$ are the dimensionless and dimensional velocity of the expansion front, respectively (Fig.~\ref{fig:fronts_transitions}a).

In these units, the UDMs are characterized by a constant dimensionless diffusion coefficient $\overline{D}=D\tau/L^2=D/D(B_0)=1$ by definition, while the VDMs exhibit a dimensionless density-dependent diffusion coefficient of the form:
\begin{equation}\label{eqn:hindered_diffusion_dimensionless}
    \overline{D}=\frac{D}{D(B_0)}=\frac{1-f(\overline{B}+\overline{I})}{1+f(\overline{B}+\overline{I})/\eta}.\frac{1+f/\eta}{1-f},
\end{equation}
where $f=B_0/B_{max}$ is the initial fraction of bacteria. Note that $\overline{D}$ corresponds to the phage diffusion coefficient relative to the diffusion coefficient at the very front of the expansion, where host density is maximal. As a consequence, $\overline{D}$ from Eq.~\ref{eqn:hindered_diffusion_dimensionless} is always greater than or equal to 1, and can therefore be interpreted as a ``boost'' in diffusion that the VDMs exhibit in the bulk of the plaque in comparison to the corresponding UDMs. This boost mathematically describes the decrease in steric interactions between phage and bacteria due to the lysis of the host as the viral infection proceeds (black line in Fig.~\ref{fig:fronts_transitions}b).

In terms of these variables, our model (Eqs. \ref{eqn:Our_Model}) becomes:
\begin{subequations}\label{eqn:Dimensionless_Model}
\begin{equation}
\pdv{\overline{B}}{\overline{t}} = -K(\beta-1)\overline{V}\,\overline{B},
\end{equation}
\begin{equation}
\pdv{\overline{I}}{\overline{t}} = K(\beta-1)\overline{V}\,\overline{B} - K(\beta-1)\overline{V}_{\overline{t}-1}\overline{B}_{\overline{t}-1},
\end{equation}
\begin{equation} 
\pdv{\overline{V}}{\overline{t}} = \pdv{}{\overline{x}} \left(\overline{D} \pdv{\overline{V}}{\overline{x}} \right) - K\overline{V}\,\overline{B} - K^*\overline{V}\overline{I} + \beta K\overline{V}_{\overline{t}-1}\overline{B}_{\overline{t}-1},
\end{equation}
\end{subequations}
where $K=\alpha\tau B_0$ and $K^*=\alpha^*\tau B_0$.

As our goal is to determine whether the travelling waves are either pulled or pushed, we will require the solution to the linearised approximation of the model. To achieve this we expand the model (Eqs.~\ref{eqn:Dimensionless_Model}) about the tip of the front where ($\overline{V}$, $\overline{B}$, $\overline{I}$) $\approx$ (0, 1, 0), keeping only linear terms. This results in the following set of equations:
\begin{subequations}\label{eqn:linearised_model}
\begin{equation}
\pdv{\overline{B}}{\overline{t}} = -K(\beta-1)\overline{V},
\end{equation}
\begin{equation}
\pdv{\overline{I}}{\overline{t}} = K(\beta-1)\overline{V} - K(\beta-1)\overline{V}_{\overline{t}-1},
\end{equation}
\begin{equation} 
\pdv{\overline{V}}{\overline{t}} = \pdv{}{\overline{x}} \left(\overline{D} \pdv{\overline{V}}{\overline{x}} \right) - K\overline{V} + \beta K\overline{V}_{\overline{t}-1}.
\end{equation}
\end{subequations}

From Eqs.~\ref{eqn:Dimensionless_Model} three natural parameters emerge: the dimensionless adsorption coefficient $K=\alpha\tau B_0$, the burst size $\beta$ and the dimensionless diffusion coefficient $\overline{D}$. In the UDMs, $\overline{D}=1$, leaving $K$ and $\beta$ as the only two parameters of the model. By contrast, in the VDMs, $\overline{D}$ is a function of $B_0$ (Eq.~\ref{eqn:hindered_diffusion_dimensionless}), which entangles the effect of initial bacterial density on $K$ and $\overline{D}$. To decouple adsorption and diffusion, we define a set of three new independent parameters that we will use in the following to analyse the model variants: the initial fraction of bacteria $f=B_0/B_{max}$, the maximum dimensionless adsorption coefficient $K_{max}=\alpha \tau B_{max}$ ($K=fK_{max}$), and the burst size $\beta$. In the linearised approximation (Eqs.~\ref{eqn:linearised_model}) $\overline{c}_F$ is the same for all four model variants and, therefore, depends only on the dimensionless adsorption coefficient $K=fK_{max}$ and the burst size $\beta$ - see Methods.

\section{From Pulled, to Semi-Pushed to Fully Pushed}
By numerically solving the PDE system in Eqs.~\ref{eqn:Dimensionless_Model}, we obtain the front velocity $\overline{c}$ and compare it with the velocity $\overline{c}_F$ of the linearised model (Eqs.~\ref{eqn:linearised_model}, see Methods for details). In addition to front velocity, we also determine the characteristic width of the infection region $\Delta\overline{x}_I$ (Fig.~\ref{fig:fronts_transitions}a), which we will discuss in Appendix~\ref{app:solid_lysis}. 

The transitions between different types of travelling wave are then determined from the ratio $\frac{\overline{c}}{\overline{c}_F}$ according to Ref.~\cite{Birzu2018}: (i) pulled waves for $\frac{\overline{c}}{\overline{c}_F}=1$, (ii) semi-pushed wave for $1<\frac{\overline{c}}{\overline{c}_F}<\frac{3}{2\sqrt{2}}$, (iii) fully pushed waves for $\frac{\overline{c}}{\overline{c}_F}\geq\frac{3}{2\sqrt{2}}$ (see Methods). We point out here that the transition between semi-pushed and fully pushed waves has been uncovered and investigated only for single species range expansions, so far. The viral model presented here is more complex because of the coupling between the dynamics of different populations (bacterial and viral) and because of the presence of a time delay. As a result, the demographic noise in our model may differ from that in Ref.~\cite{Birzu2018}. While this does not affect the transition between pulled and pushed waves, the distinction between semi-pushed and fully pushed waves might, in principle, be different. 

The location of these transitions in the different model variants for a set of infection parameters typical of T7 is shown in Fig.~\ref{fig:fronts_transitions}c. Under these conditions, we observe that the UDM+ exhibits a pulled wave for the full range of initial bacterial fraction, while the UDM-, the VDM+ and the VDM- waves become increasingly more pushed as $f$ increases. In terms of dimensional velocity, the difference between the model variants is minimal (inset in Fig.~\ref{fig:fronts_transitions}c), justifying why these effects have gone unnoticed in past theoretical work that aimed at predicting experimental phage front speeds. 

\begin{figure}
\includegraphics[width=0.5\textwidth]{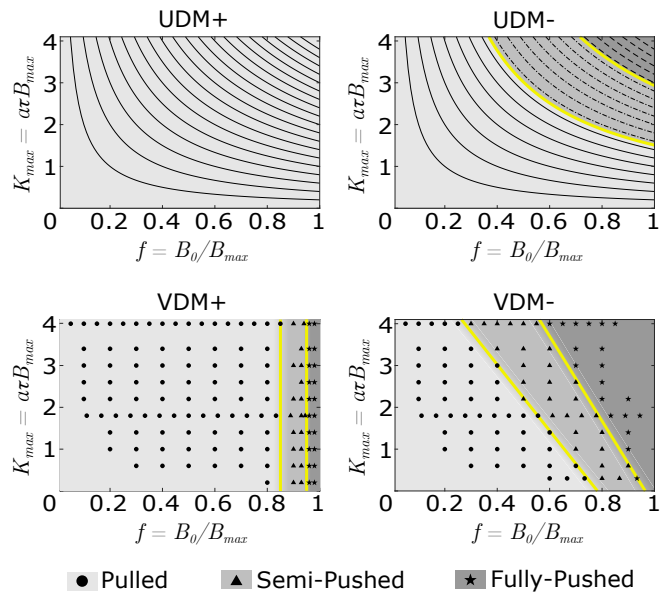}
\caption{\label{fig:phase} Phase diagrams showing the expansion types for the four model variants as a function of bacterial density $f$ and maximum dimensionless adsorption coefficient $K_{max}$ - burst size $\beta$=50 throughout. Lines in the UDMs, and data points in the VDMs indicate the parameter combinations for which numerical integration was performed, and velocities calculated. These values are interpolated to estimate the transition boundaries between different classes of travelling waves (yellow lines). In the UDM+, we do not observe pushed waves, while in the VDM+ transitions between pulled and pushed waves occur at approximately constant bacteria fractions. In the UDM-, as $K$ and $\beta$ are the only free parameters, transitions occur at specific values of $K$.  In the VDM-, the transitions are heuristically approximated as linear relationships with gradient $m$ and intercept $a$ ($f=mK_{max}+a$).} 
\end{figure}

\subsection{Wave Transitions are Very Sensitive to Virus-Host Interactions}
To generalise our findings and fully characterise the origin and nature of the transitions in front dynamic for the different model variants, we extend our investigation to a broader range of parameter values, by varying $K_{max}$ (and $\beta$, see Appendix~\ref{app:burst}) about the parameters used in Fig.~\ref{fig:fronts_transitions}c. 

\begin{figure*}
\includegraphics[width=0.99\textwidth]{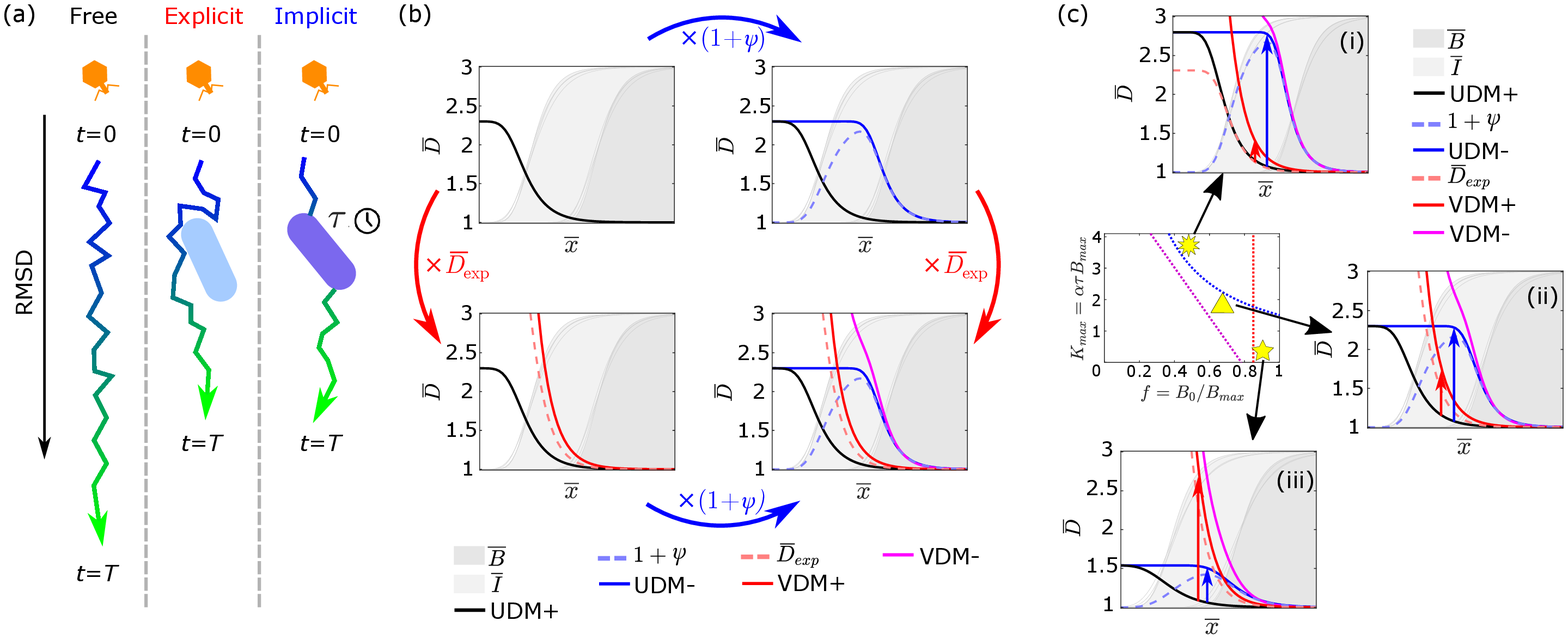}
\caption{\label{fig:implicit_diffusion} (a): An illustration of both explicit and implicit effects to phage diffusion. Due to the explicit effect, phage diffusion is hindered by steric interactions with bacterial hosts, while the implicit effect hinders phage diffusion by trapping the virus for a period $\tau$ during which it cannot disperse. Colour gradients on the phage trajectories indicate the passage of time. (b): Proxies for the diffusive behaviour in each of the model variants plotted as a function of position across the expansion front. The base diffusion coefficient $\overline{D}_{imp+}$ (Eq.~\ref{eqn:Dimp+}) in the UDM+ (black solid line) is modified either by the term $1+\psi$ (blue arrow and blue dashed line) in the UDM- (blue solid line), which accounts for the now unhindered diffusion in the region of infected cells, or by an additional term $\overline{D}_{exp}$ (red arrow and red dashed line) in the VDM+ (red solid line) which accounts for the hindrance due to steric effects. Both modifications occur in the VDM- (magenta line). Faint grey lines indicate the different front profiles from the model variants used to calculate the diffusion coefficient profiles (see Methods). (c): Diffusion profiles for three representative cases are shown, all highlighted in comparison to the semi-pushed transition lines for the models shown in Fig.~\ref{fig:phase}: (i) high $K_{max}$ and low $f$, where the UDM- and VDM- expansions are pushed; (ii) intermediate $f$ and $K_{max}$, where only the VDM- expansion is pushed; (iii) low $K_{max}$ and high $f$, where the VDM+ and VDM- expansions are pushed. Red and blue arrows highlight the shift in $\overline{D}$ from the UDM+ to the VDM+ and UDM- respectively at the position where the viral profile is approximately 3/4 times its steady-state.}
\end{figure*}

Fig.~\ref{fig:phase} shows the type of expansion that occurs in each of the models as a function of $f$ and $K_{max}$. The results clearly indicate that the presence or absence of density-dependent diffusion and adsorption to infected cells can dramatically alter the type of travelling wave undergone by phage, with the UDM+ being the only model resulting in a pulled population wave for the whole range of parameters explored. 

In the following, we will provide a physical interpretation for these observations, by identifying two independent mechanisms that alter phage dispersal in a density-dependent fashion. The first (Sec.~\ref{sec:explicit}), which we name the `explicit' effect, is caused by steric interactions between phage and the bacterial host, and represents the effect measured in our experiments (Fig.~\ref{fig:Diffusion_Experiment}). The second (Sec.~\ref{sec:implicit}), which we name the `implicit' effect, arises spontaneously from the infection dynamic due to the fact that during incubation, phage are \textit{trapped} inside the host cells, unable to diffuse, thereby resulting in a density-dependent effect on phage diffusion.

\subsection{Decreased Steric Effects due to Host Lysis Promote the Transition to Pushed Waves at High Bacterial Densities}\label{sec:explicit}
The effect of virus-host steric interactions can be best appreciated by comparing the phase diagram of the UDM+ to that of the VDM+, and is a direct consequence of the variable diffusion coefficient that our model \textit{explicitly} introduces in Eq.~\ref{eqn:hindered_diffusion_dimensionless}.

In the VDM+, transitions to semi-pushed and fully pushed waves occur at high values of $f$ with very weak dependence on $K_{max}$. This results from the boost in phage diffusion that occurs in the bulk of the plaque as host cells lyse and steric effects decrease. Because the boost increases with increasing difference in bacterial density between the front ($\overline{B}=B/B_0=1$) and the back ($\overline{B}=0$), higher initial bacterial density $f$ will lead to a stronger boost. We find empirically that, beyond a given point controlled exclusively by $f$, the phage behind the propagating front will disperse sufficiently fast to be able to catch up with the front and generate a semi-pushed or even a fully-pushed wave. For what follows, it is useful to name this explicit boost to diffusion $\overline{D}_{exp}$, which is mathematically identical to the dimensionless diffusion coefficient in Eq.~\ref{eqn:hindered_diffusion_dimensionless}, and reaches its maximum in the bulk of the plaque where no bacteria are left (Fig.~\ref{fig:fronts_transitions}b and dashed red line in Fig.~\ref{fig:implicit_diffusion}b).

\subsection{A Second ``Implicit'' Density-Dependent Diffusion Emerges from the Viral Infection Dynamics}\label{sec:implicit}
Since the UDMs lack the explicit density-dependent diffusion, the appearance of transitions to pushed regimes in the UDM- may seem surprising (Fig.~\ref{fig:phase}). To understand the origin of these transitions, it is helpful to consider the effects of the parameter $K=\alpha \tau B_0$ that controls the transition. Adsorption and incubation (quantified by the parameter $K$) are not only key for the effective growth rate of the phage population, but also for the effective dispersal rate of the phage, as they control the time and the probability that phage particles are ``trapped" in a host cell, unable to disperse. As $K$ increases, either \textit{more} phage adsorb to host cells per unit of time (increased adsorption rate), or they are trapped in the host for \textit{longer} (increased lysis time), resulting in a hampered dispersal of the phage (Fig.~\ref{fig:implicit_diffusion}a). The strength of this effect, by which phage is kept \textit{prisoner} by the host cell, has to depend on the number of host available to infect, and thus be the strongest at the edge of the expansion, where there is plenty of uninfected host, and the weakest in the bulk, where all the host has been removed. Beyond a certain point, we therefore expect phage diffusion to be sufficiently hindered at the front to allow the phage in the back to catch up and generate a pushed wave.

To quantify the reduced dispersal resulting from viral infection, we consider a system of point-like phage particles diffusing across a field of completely permeable ``sticky'' obstacles, mimicking host bacteria that trap phage for a time $\tau$. A simple mean-field analytical argument (see Methods), supported by two-dimensional Monte Carlo simulations (Appendix~\ref{app:MonteCarlo}), demonstrates that the particles in this system exhibit a hindered diffusion $D$ compared to their free diffusion $D_0=D(B=0)$, such that
\begin{equation}\label{eqn:implicit_diffusion}
    \frac{D}{D_0}=\hat{D}_{imp}=\frac{1}{1+bK_{max}},
\end{equation}
where $b$ is the local density of host that can be infected by phage relative to the density $B_{max}$ at which diffusion is completely prevented. We note here that, by definition, $\hat{D}$ is the phage diffusion coefficient relative to the bulk of the expansion, in parallel to $\overline{D}=D/D(B_0)$, defined earlier, which is the phage diffusion coefficient relative to the front of the expansion.

When adsorption to infected cells occurs (UDM+), infected cells \textit{trap} phage as much as uninfected cells ($b=(B+I)/B_{max}$ in Eq.~\ref{eqn:implicit_diffusion}) resulting in a diffusion coefficient in the dimensionless model of the form
\begin{equation}\label{eqn:Dimp+}
    \overline{D}_{imp+}=\frac{\hat{D}_{imp+}}{\hat{D}_{imp+}(B_0)}=\frac{1+K}{1+(\overline{B}+\overline{I})K}.
\end{equation}
Because of the shape of the bacterial density profile during the expansion, phage diffusion will then be highest in the bulk and slowest at the front (black line in Fig.~\ref{fig:implicit_diffusion}b). Yet, fast diffusing phage appear too far from the front to contribute to the expansion, resulting in pulled waves across parameter space (Fig.~\ref{fig:phase}).

By contrast, when adsorption to infected cells is prevented (UDM-), phage can no longer become trapped in the infected region behind the front ($b=B/B_{max}$ in Eq.~\ref{eqn:implicit_diffusion}), so that
\begin{equation}
\overline{D}_{imp-}=\frac{\hat{D}_{imp-}}{\hat{D}_{imp-}(B_0)}=\frac{1+K}{1+\overline{B}K},
\end{equation}
and fast diffusing phage emerge much closer to the expansion front (blue vs. black lines in Fig.~\ref{fig:implicit_diffusion}b, Methods). Preventing adsorption to infected cells is therefore equivalent to a boost in implicit diffusion in the infected region just behind the front, which can be approximated to 
\begin{equation}
    \frac{\overline{D}_{imp-}}{\overline{D}_{imp+}}=1+\psi\approx 1+\frac{\overline{I}K}{1+\overline{B}K},
\end{equation}
(blue dashed line in Fig.~\ref{fig:implicit_diffusion}b). This boost is sufficient to shift the fast diffusing phage closer to the expanding front and, if $K$ is sufficiently large, to generate a transition to pushed waves.

It is important to point out that this implicit density-dependent diffusion emerges spontaneously from the viral infection dynamics (common to most viruses), where infecting viruses trapped in the host cannot contribute to the advancement of the front until they are released from the host. As a consequence, unlike the explicit density-dependent diffusion, this effect cannot be easily accommodated into the diffusion coefficient of our model, as it does not act independently of the infection and growth processes. Indeed, an alternative interpretation for this mechanism can be provided by a density-dependent death rate. In our model, adsorption to previously infected hosts is equivalent to phage `death', as it results in the permanent loss of these phage. Going from the case where adsorption to infected cells occurs to the case where it does not (from + to - models) will then lead to an increase in net growth rate in the region of infected cells, which lies at intermediate viral density (Fig.~\ref{fig:fronts_transitions}a). The result is a higher net growth rate at intermediate population densities similar to what an Allee effect would generate in a mono-species expansion~\cite{Allee1931, Roques2012}.

\subsection{Implicit and Explicit Density-Dependent Diffusions Act Independently with Multiplicative Effects}
Because the implicit and explicit boosts to diffusion discussed above have different physical origins and are controlled by different parameters ($K$ and $f$, respectively), they play significant roles in different regions of parameter space. The implicit boost that results from a lack of adsorption to infected cells, encoded in ($1+\psi$), is stronger at large $K_{max}$, where more phages are trapped by hosts for a longer period of time. Instead, the explicit boost caused by steric interactions, encoded in $\overline{D}_{exp}$, is dominant at low $K_{max}$. The ratio of the two effects over parameter space is shown in Appendix~\ref{app:effects_ratio}, Fig.~\ref{fig:effects_ratio}.

Extending the analytical argument with which we defined the implicit boost to diffusion, we can show that, to a first approximation, explicit and implicit effects act independently over a basal diffusion coefficient (see Methods). As a consequence, preventing adsorption to infected cells corresponds to multiplying the diffusion coefficient by $1+\psi$ (from + to - models, blue arrows in Fig.~\ref{fig:implicit_diffusion}). Similarly, including steric effects corresponds to multiplying the diffusion coefficient by $\overline{D}_{exp}$ (from UD to VD models, red arrows in Fig.~\ref{fig:implicit_diffusion}). As a result, we can write the dimensionless diffusion coefficient of the VDM-, which exhibits both effects, as
\begin{equation}\label{eqn:VDM-}
    \overline{D}_{VDM-}=(1+\psi)\overline{D}_{exp}\overline{D}_{imp+},
\end{equation}
where all the terms are calculated with respect to $\overline{B}$ and $\overline{I}$ from the VDM- simulations. 

\begin{figure}[h]
\includegraphics[width=0.5\textwidth]{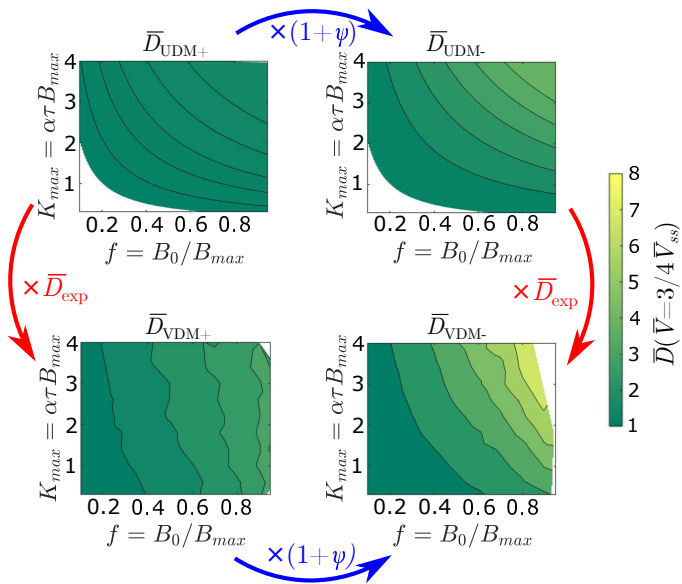}
\caption{\label{fig:implicit_phase} Dimensionless diffusion coefficients in each of the models determined at the front position where the phage population is 3/4 times the steady state population $\overline{V}_{ss}$, plotted as a function of both $f$ and $K_{max}$. Contour lines indicate levels of constant $\overline{D}$. The behaviour of the contours qualitatively matches that of the transition boundaries in Fig.~\ref{fig:phase}.}
\end{figure}

In contrast to the other models, this function depends non-trivially on $K_{max}$ and $f$, making it challenging to find a simple parameter combination that controls the transitions to pushed waves. Nonetheless, we see that the diffusion coefficient determined at 3/4 times the steady state phage population is able to qualitatively capture the behavior of the transition lines in all models (Fig.~\ref{fig:implicit_phase}) and it explains why the transition lines in the VDM- approaches the transition lines in the UDM- and the VDM+ at high and low $K_{max}$, respectively, where either effects dominate (Fig.~\ref{fig:phase} and Fig.~\ref{fig:implicit_diffusion}c). While the phage diffusion at a specific population density is, in principle, insufficient to predict whether the expansion is pushed, which by definition depends on the whole wave dynamic, Fig.~\ref{fig:implicit_phase} illustrates that regions in parameter space with similar effective diffusion within a model correspond to similar types of expansions. This supports the idea that the density-dependent diffusion, whether implicit or explicit, is the key ingredient that leads to transitions to pushed waves. 

Further analyses indicate that our results are robust to (i) changes in burst size (Appendix~\ref{app:burst}), (ii) inclusion of bacterial growth (Appendix~\ref{app:growth}) and (iii) phage adsorption to cell debris in the plaque centre (Appendix~\ref{app:debris}). In short, we find that changes in burst size do not qualitatively change the shape of the transitions (Fig.~\ref{fig:phase_Y20}), but slightly alter their location (Fig.~\ref{fig:burst}), with lower burst sizes facilitating the transition to pushed waves. We also find that the introduction of bacteria growth at a rate consistent with \textit{E. coli} has no discernible effect (Fig.~\ref{fig:growth}). Finally, the inclusion of adsorbing cell debris does not qualitatively change the behaviour of the model, but can narrow the region of parameter space where pushed waves are expected (Fig.~\ref{fig:debris}a and b). This effect is, however, very limited for rates of adsorption to debris that can be considered physically realistic (Fig.~\ref{fig:debris2}).

\subsection{Lysis Time is the Main Determinant for the Rate of Genetic Diversity Loss}
\begin{figure*}
\includegraphics[width=0.99\textwidth]{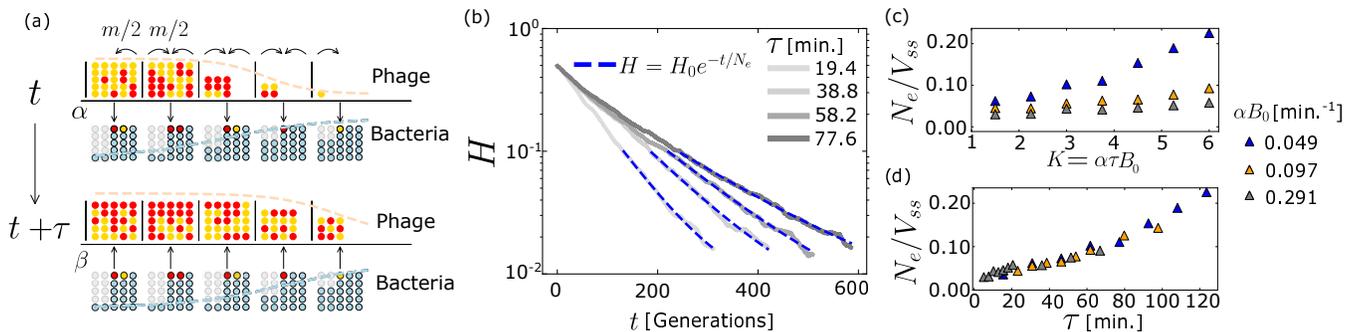}
\caption{\label{fig:het_data} (a): Simulation algorithm of stochastic simulations of plaque growth.
Each deme contains two labelled phage populations (yellow and red circles) and bacteria that can be dead (grey), uninfected (light blue) or infected with one type of phage (red/yellow). At each step, phage can migrate to neighboring demes with probability m/2  (right/left curved arrows), infect an uninfected cell with probability $ \alpha $ (downwards arrow), and after $\tau$ time steps, introduce $ \beta$ phage into the deme upon cell lysis (upwards arrow). Dashed lines indicate the analogous density profile, similar to those shown in Fig.~\ref{fig:fronts_transitions}a. (b): Example of linear fit to log transformed heterozygosity data, shown for $\alpha B_0 = 0.077$ min$^{-1}$, $B_0=100$ cells/deme, and a range of $\tau$ values. Heterozygosity data represents the average of $\sim$1000 simulations. Slope of fit, $1 / N_e$, is the calculated decay rate for the given parameter values. (c,d): Effective population size $N_e$ for $B_0 = 75$, normalized by the steady state population in the bulk of the population $V_{ss}$ as a function of $K$ (c) and $\tau$ (d), over a range that could be expected in various viral populations. Colors indicate specific values of $\alpha B_0$. Errors in (c) and (d) due to linear fit of heterozygosity decay over time are negligibly small and not shown.}
\end{figure*}

Single-species populations expanding via pushed waves have been theoretically and experimentally shown to retain genetic diversity much longer than their pulled counterparts \cite{Hallatschek2008,Roques2012, Birzu2018, Gandhi2019}. To test whether this property is maintained in viral populations and determine the effect of virus-host interactions on the rate of diversity loss, we developed an agent-based stochastic implementation of one of our numerical models and tracked the viral heterozygosity $H$ as a function of time~(Fig.~\ref{fig:het_data}, Methods). The heterozygosity $H$ in a viral biallelic population is given by
\begin{equation}
H = \frac{1}{M}\sum^M_i 2f_i(1-f_i),
\end{equation}
where $M$ is the total number of demes in the simulation box, and the fraction of the two alleles in deme $i$ are $f_i$ and $1-f_i$. We focus on the UDM- as it is the simplest of our models that exhibits pushed waves, and it is also relevant for viruses beyond phage T7. 

Analogously to previous studies on single-species, we find that the heterozygosity decays exponentially over time, so that we can define an effective population size $N_e$ as the inverse of the decay rate in units of generations (Fig.~\ref{fig:het_data}b,  see Methods for details). To account for the fact that adsorption rate, lysis time and bacterial density also change the density profile of the viral population, we normalize the effective population size by the steady-state value of the viral population in the bulk of the wave $V_{ss}$. This normalization aims at providing a direct comparison between our system and previous theoretical studies where the carrying capacity of the population was held constant (see Appendix~\ref{app:Het_extra} for data without normalization). 

Our results show that the level of $\textit{pushedness}$ of the wave, controlled by $K$ in the UDM-, can signficantly increase the normalized effective population size (two-fold increase between a pulled wave, $K=1.5$, and just above the pushed transition, $K=3$, Fig.~\ref{fig:het_data}c). However, we also find that, not surprisingly, $K$ alone is not sufficient to determine the value of the effective population size of the expansion. Similar observations have been made before in single-species expansions, where distinct cooperativity models, all displaying transitions between pulled and pushed waves, were found to be characterised by different values of $N_e$~\cite{Birzu2019}.
Remarkably, we find that an excellent predictor for the value of $N_e$ in the UDM- is the lysis time $\tau$ (collapse of datasets in Fig.~\ref{fig:het_data}d and Fig.~\ref{fig:het_supp}). A possible reason behind this observation is the different effects that lysis time and adsorption rate have on the steady-state viral density profile (one broadens it, while the other narrows it), which, in turn, impacts $N_e$~\cite{Hallatschek2008,Birzu2018,Birzu2019}. Further analyses are necessary to pin-point the exact mechanisms that link virus-host interactions and viral diversity, urging for future theoretical work to investigate viral genetic diversity in spatial settings.

\section{Discussion}
In this work, we first experimentally quantify how the diffusion of phage in a bacterial lawn is hindered by steric interactions with the host bacterial cells, resulting in a density-dependent diffusion coefficient. Going beyond current descriptions of plaque growth, which have considered host density-dependence only for setting a constant diffusion coefficient parameter, we construct a reaction-diffusion model of the phage-bacteria system that explicitly incorporates a diffusion coefficient that depends on local host density, and therefore varies in time and space. We show that, in contrast to current thinking which assumes that viral expansions are always pulled, this `explicit' effect can lead to a transition from pulled to pushed waves at high host densities. We also show that a second, independent density-dependence in diffusion emerges implicitly from the underlying viral dynamics, whereby phage are unable to disperse during replication within the host. We find that when adsorption to infected host cells is prevented, this `implicit' effect can also lead to the transition to pushed waves. Together, this indicates that bacteriophage offer an excellent experimental system to study the effect of density-dependent diffusion on expansion dynamics.

The transition from a pulled wave to a pushed wave has traditionally been associated with increased co-operativity between individuals, quantified by density-dependent growth, or more recently, density-dependent dispersal \cite{Allee1931, Birzu2018, Birzu2019}. By analogy, the density-dependence in phage diffusion can be interpreted as an emergent co-operativity, which stems from the fact that as phage work together towards cell lysis, they remove bacterial obstacles, indirectly favouring the dispersal of neighboring phage. The fact that the diffusion is dynamically changed as phage replicate could lead to interesting ecological feedback. Ecological feedback on diffusion has been theoretically shown in other contexts to lead to pattern formation, and in some cases help maintain genetic diversity and mitigate the risk of extinctions \cite{Park2019}. Indeed, density-dependent dispersal has been identified as a key ingredient in a generic route to pattern formation in bacterial populations \cite{Cates2010}.

We find that the transition to a pushed wave can occur due to two separate effects: an explicit density-dependent diffusion coefficient, caused by steric interactions between the phage and the host bacteria, which is dominant in crowded host environments, and an implicit hindrance to the diffusion of the phage population at the front caused by the viral infection dynamics. We therefore expect that the pushed dynamics will be strongest in populations that experience both effects, and where adsorption to infected host is absent (VDM-). Some bacteriophage have mechanisms that prevent adsorption to already infected cells, usually by blocking receptor sites post-infection \cite{Labrie2010}. Bacteriophage T5 produces a lipoprotein (Llp) that is expressed at the beginning of infection, preventing superinfection by blocking its own receptor site (FhuA protein), and protecting newly produced phage from inactivation by binding to free receptors released by lysed cells \cite{Braun1994, Perduzzi1998}. Similar mechanisms are also well documented in several temperate phage. Phage $\Phi$V10 possesses an O-acetyltransferase that modifies the specific $\Phi$V10 receptor site (the O-antigen of \textit{E. coli} O157:H7) to block adsorption \cite{Perry2009}. Similarly, \textit{Pseudomonas aeruginosa} prophage D3 modifies the O-antigen of LPS on the host surface to prevent adsorption of the many phage that bind to the O-antigen \cite{Newton2004}. This is similar to other \textit{Pseudomonas} prophage which encode for twitching-inhibitory protein (TiP) that modifies the type IV pilus on the \textit{P. aeruginosa}, preventing further adsorption \cite{BondyDenomy2016, vanHoute2016}.

Mechanisms that prevent superinfection by forbidding adsorption to infected cells have been observed in viruses beyond bacteriophage \cite{Walsh2019}. For instance, cells recently infected with Vaccinia virus VacV (the live vaccine used to eradicate smallpox) express two proteins that repel super-infecting virions, resulting in plaques that grow four-fold faster than predicted by replication kinetics alone \cite{Doceul2010}. Our results show that even in the absence of explicit steric effects, pushed expansions can occur if adsorption to infected hosts is prevented (UDM-), as is the case for VacV, simply due to the fact that viruses are unable to disperse during incubation, suggesting that pushed waves might be far more widespread than previously thought among different viral systems.

Pushed dynamics in range expansions have been shown to have significant consequences for the evolution of the population. In pulled expansions, the high susceptibility to stochastic fluctuations results in inefficient selection, as beneficial or deleterious mutations can effectively behave as neutral due to the small number of individuals contributing to the dynamics \cite{Panja2004, Matsuyama1994, Giometto2014}, and leading, for instance, to the accumulation of deleterious mutations, known as expansion load \cite{Peischl2015, Bosshard2017}. In fully pushed waves, stochastic fluctuations are much weaker as more individuals contribute to the advancement of the population, allowing beneficial mutations to establish more easily and deleterious mutations to be purged~\cite{Roques2012,Birzu2019}. Our stochastic simulations show that even the implicit density-dependent diffusion alone can slow down the rate of diversity loss up to 5-fold under reasonable phage infection parameters (Fig.~\ref{fig:het_data}). Remarkably, we find that the rate of diversity loss strongly depends on the lysis time, but only weakly on adsorption ($\alpha B_0$), even if the two parameters are expected to contribute equally to the level of ``pushedness'' of the wave. This observation reveals a rich and non-trivial evolutionary dynamic displayed by our viral model that distinguishes it from classic mathematical descriptions of pushed waves, where dispersal, growth and cooperativity are controlled by independent parameters.

Going forward, three clear avenues emerge as a result of our work. Firstly, the complex dependence of the expansion dynamics on the infection parameters that we observe indicates that viral expansions offer a currently untapped ground for further theoretical studies. Our model provides a framework to investigate the evolutionary dynamics of an expanding viral population in terms of the realistic processes that occur therein. Within this framework, future work is required to fully characterise the complex interplay that each of the infection processes exhibit, and ultimately determine what impact they have on the viral evolutionary dynamics.

Secondly, while this work provides theoretical predictions and physical insights regarding the transition from pulled to pushed waves in viral expansions, it also points at phage plaques as a well-controlled model system to experimentally investigate these theories in a laboratory setting. We have shown (Fig.~\ref{fig:fronts_transitions}c and Fig.~\ref{fig:phase}) that pushed waves can occur during plaque growth in conditions easily achievable in the laboratory. While it is challenging in a laboratory setting to fully replicate the complex environments found in nature, we believe that plaques offer an alternative and possibly more realistic environment to study these dynamics than the environments typically used thus far, such as cultures in a 96-well plate, where dispersal is achieved by artificial migration schemes~\cite{Kerr2006, Gandhi2016, Gandhi2019}.

Lastly, our experiments have shown that the rate of phage diffusion strongly depends on the host environment and, in particular, can dramatically differ from liquid culture measurements, where even at high overnight densities of $\sim10^{9}$ cells/ml, the volume fraction occupied by the cells is $\sim0.001$, and so diffusion is effectively unhindered. This realisation raises the more general question of whether other phage life history parameters, such as lysis time (Appendix~\ref{app:solid_lysis}), also depend strongly on the surrounding host environment. Traditionally, these parameters are determined by experiments carried out in a well mixed liquid culture \cite{Yin1993, Fort2002}. It is perfectly possible that these parameters could take significantly different values when the infection occurs in different spatially structured environments and varying metabolic states of the host cells \cite{Choua2019}. Moreover, it is possible that these parameters are not only different on solid media when compared to liquid, but may also vary across the expansion in a similar fashion to the diffusion coefficient. For instance, upon lysis, release of cytoplasmic fluids could affect the infection or lysis of neighbouring cells, resulting in life-history parameters that vary with cell density.

\section{Materials and Methods}\label{sec:methods}
\subsection*{Bacterial Strains} 
Five strains of \textit{E. coli} were involved in this work. The first strain, \textit{E. coli} BW25113 (CGSC\# 7636), is susceptible to phage infection. This strain was transformed previously with a plasmid expressing Venus YFP to create the second strain, \textit{E. coli} eWM43 \cite{Mobius2015}. This strain was further transformed with plasmid pAK501 (Addgene\# 48107) \cite{Kaczmarczyk2013}, which confers resistance to chloramphenicol, to create the third strain, \textit{E. coli} eMTH43. The fourth strain used, \textit{E. coli $\Delta$waaC}, is resistant to phage infection through deletion of the \textit{waaC} gene, the product of which is involved in the production of lipopolysaccharide, the recognition of which is essential for the adsorption of phage \cite{Qimron2006}. This strain was transformed previously with a plasmid expressing mCherry to yield the final strain \textit{E. coli} eWM44 \cite{Mobius2015}. The strains eMTH43 and eWM44 were the two strains used respectively as the susceptible and resistant host in our experiments.

\subsection*{Bacteriophage T7} 
The phage used in the study is the obligately lytic bacteriophage T7. The phage was originally obtained as an aliquot from the wild-type stock of the Richardson lab (Harvard Medical School, Boston, MA). To prepare stocks of this phage, phage were added to an exponentially growing liquid culture of BW25113, and incubated at 37 \degree C until clear. The lysate was then mixed with NaCl to a final concentration of 1.4 M. This was then spun down to remove cell debris, and the resulting supernatant was stored at 4 \degree C.

\subsection*{Sample Preparation}
To measure the diffusion coefficient of phage, 96 well plate sized omni-plates, containing 35 ml of 20 g/l agar (VWR Chemicals), with LB (Invitrogen) - NaCl concentration 10 g/l - and 15 $\mu$g/ml  chloramphenicol (Sigma-Aldrich) were prepared and kept at room temperature for 2 days. The presence of chloramphenicol in the plate prevents growth of the background strain eWM44, so to maintain its density constant over the course of the experiment. Plates were then refrigerated if they were to be used at a later date. Overnight liquid cultures of \textit{E. coli} were grown from single colonies at 37 \degree C in LB with either 100 $\mu$g/ml ampicillin (Sigma-Aldrich) or 15 $\mu$g/ml chloramphenicol (Sigma-Aldrich) for eWM44 and eMTH43 respectively. 

To create the background lawn of bacteria, the  optical density at 600 nm (OD$_{600}$) of the eWM44 culture was measured, and diluted into LB to obtain the desired density (calculated on the basis that OD$_{600}=0.1$ equates to $10^{8}$ cells/ml). A 500 $\mu$l droplet of this culture was then spread with glass beads (radius 4 mm) across the surface of the agar until dry. This process (a 500 $\mu$l droplet spread with fresh beads) was repeated a further two times to achieve as uniform a distribution as possible. The plate was then left for a further 10 minutes before proceeding to the next step.

10 ml of eMTH43 overnight culture was spun down and re-suspended in fresh LB, so as to give an OD$_{600}$ reading of 0.50 if diluted hundredfold. 2 $\mu$l droplets of the concentrated culture were then pipetted onto the lawn of eWM44 in a grid like pattern, spaced approximately 1 cm apart, and left to dry (approximately 15 mins after the last droplet was pipetted). Each plate contained approximately 60 host droplets. The plate was then incubated at 37 \degree C for 1 hour.

10 $\mu$l of stock T7 phage (10$^7$ pfu/ml) was diluted in 100 $\mu$l of black food dye. 1 $\mu$l droplets of this dilution were pipetted onto the surface of the agar, in the gaps between the previously pipetted droplets of eMTH43, and left to dry. A schematic of the resulting set-up can be seen in Fig.~\ref{fig:Diffusion_Experiment}a.

\subsection*{Data Acquisition}
The plates were imaged using a Zeiss Axio Zoom.V16 stereo microscope equipped with a Zeiss PlanApo Z 0.5x/0.125 FWD 114 mm objective. Images of the sample were taken every 20 minutes for a period of 24 hours. During the imaging period, the sample was kept with its lid on at 37 \degree C using an ibidi Multi-Well Plate Heating System.

\subsection*{Data Analysis}
All of the images were analysed using Fiji (v1.52h), an open source distribution of ImageJ focused on scientific image analysis \cite{Schindelin2012, Rueden2017}. 

The time $\Delta t$ necessary for a clearing to appear in the droplets of eMTH43, and the separation $r$ between the point at which a clearing forms and the nearest phage droplet was recorded (Fig.~\ref{fig:Diffusion_Experiment}). By measuring $r$ and $\Delta t$ over many droplet-droplet pairs, one can measure the mean first-passage time $\langle t \rangle$ (the mean time taken for the first phage to diffuse a fixed distance $r$), and calculate the rate of phage diffusion by fitting the relationship \cite{Klein1952}:
\begin{equation}
     r^2 = 4D \langle t \rangle + \textrm{constant},
\end{equation}
where $D$ is the diffusion coefficient, and the constant arises due to the delay between the arrival of the phage and the formation of a visible plaque.

\subsection*{Numerical Solutions of Reaction-Diffusion Model}
We consider the scaled equations (Eq. \ref{eqn:Dimensionless_Model}) on a finite interval of length $L_D$ with homogeneous Neumann boundary conditions. Throughout we used $L_D=120$ and a maximum possible time of $\overline{t}_{max}=50$. Initially, we set $\overline{B}=1$ over the whole interval, $\overline{V}=1$ for $\overline{x}\leq2$, and $\overline{V}=0$ elsewhere. There are initially no infected bacteria ($\overline{I}=0$). Solutions are determined on a mesh of uniform space and time, with divisions of $d\overline{t}=0.1$, and $d\overline{x}=0.1$ or $d\overline{x}=0.2$ to give the best balance between precision and compute time for a given parameter set.

A sketch of the fronts during the expansion can be seen in Fig.~\ref{fig:fronts_transitions}a. The dimensionless spreading velocity $\overline{c}$ of the front is determined by tracking the midpoint of the bacterial wave (i.e. $\overline{B}=0.5$) over time. For pulled fronts, the spreading velocity is known to demonstrate a power law convergence to an asymptotic value \cite{Ebert2000}. In the case where a steady spreading velocity was not reached, the spreading velocity was given by the asymptotic value which produced the best power law fit to the data.

The transitions between pulled, semi-pushed and fully pushed have been found to occur at specific wave velocities with respect to the linearised (Fisher) velocity $c_{F}$ - the velocity determined solely by the linear dynamics at the tip of the front \cite{Birzu2018}. Pulled expansions spread with a velocity equal to the velocity of the linearised model $c=c_{F}$, while pushed expansions spread faster \cite{Saarloos2003}. The transition between semi-pushed and fully pushed occurs at a velocity of $\frac{c}{c_{F}}=\frac{3}{2\sqrt{2}}$, with waves below this velocity being semi-pushed, and above this velocity being fully pushed \cite{Birzu2018}. These thresholds have been shown to be robust to the details of the population dynamics~\cite{Birzu2019}, though their appropriateness for multi-species expansions requires further investigations. We here use the same values for illustration purposes.

Even though strictly speaking pulled expansions occur when the spreading velocity is equal to the velocity of the linearised model ($c=c_{F}$), due to errors in determining the velocity over a finite time (i.e. errors due to the power law fit when determining the asymptotic velocity), we conservatively consider velocities within 1\% of the linearised velocity as corresponding to pulled expansions.

A length scale characterising the width of the infected region $\Delta\overline{x}_I$ was also computed by tracking the separation between the midpoint on the wave of uninfected bacteria ($\overline{B}=0.5$), and the midpoint on the wave of infected bacteria ($\overline{B}+\overline{I}=0.5$) over time. An average was taken over the final 20 time points for the reported value of $\Delta\overline{x}_I$.

\subsection*{Linearised Solution of Reaction-Diffusion Model}
To determine the transition between pulled, semi-pushed and fully pushed regimes, the solution to the linearised model is required. To achieve this, we first look for travelling wave solutions to Eq. \ref{eqn:Dimensionless_Model} in the co-moving coordinate $\overline{z}\equiv \overline{x}-\overline{c}\overline{t}$ where $\overline{c}$ is the dimensionless front velocity.

As the components approach their limiting concentrations at the leading edge of the front, the linearised form of the model becomes valid, and so, following previous work \cite{Yin1992, Fort2002}, we assume the concentrations take the form $\overline{V}=a_1 \textrm{exp}(-\overline{\lambda}\overline{z})$, $\overline{B}=1-a_2 \textrm{exp}(-\overline{\lambda}\overline{z})$ and $\overline{I}=a_3 \textrm{exp}(-\overline{\lambda}\overline{z})$ where $\overline{\lambda}$ is a dimensionless width parameter, and $a_1$, $a_2$ and $a_3$ are positive constants.

Substituting into the linearised version of the model (Eqs.~\ref{eqn:linearised_model}) and writing in matrix notation yields:
\begin{equation}
    \begin{pmatrix}
    K(\beta-1) & -\overline{\lambda}\overline{c} & 0 \\
    K(\beta-1)(1-\textrm{e}^{-\overline{\lambda}\overline{c}}) & 0 & -\overline{\lambda}\overline{c} \\
    \overline{\lambda}^2 - \overline{\lambda}\overline{c} + K(\beta\textrm{e}^{-\overline{\lambda}\overline{c}}-1) & 0 & 0
    \end{pmatrix}
    \begin{pmatrix}
    a_1 \\
    a_2 \\
    a_3
    \end{pmatrix}
    =0.
\end{equation}
To find non-trivial solutions the determinant of the matrix is set to zero, leading to the characteristic equation:
\begin{equation}\label{eqn:characteristic}
    \overline{\lambda}^2 - \overline{\lambda}\overline{c} + K(\beta\textrm{e}^{-\overline{\lambda}\overline{c}}-1)=0.
\end{equation}

As we are assuming here that the front is pulled, the front propagates with the minimum possible velocity \cite{Ebert2000}:
\begin{equation}\label{eqn:dcdl0}
    \overline{c} = \textrm{min}_{\overline{\lambda}>0}[\overline{c}(\overline{\lambda})].  
\end{equation}
By implicitly differentiating Eq.~\ref{eqn:characteristic} with respect to $\overline{\lambda}$, and setting $d\overline{c}/d\overline{\lambda}=0$ according to Eq.~\ref{eqn:dcdl0}, this leads to the second characteristic equation:
\begin{equation}\label{eqn:characteristic2}
    2\overline{\lambda}-\overline{c}-K\beta\overline{c}\textrm{e}^{-\overline{\lambda}\overline{c}}=0.
\end{equation}
The dimensionless spreading velocity $\overline{c}$ is given as the unique solution to both Eq. \ref{eqn:characteristic} and Eq. \ref{eqn:characteristic2} which we solved numerically for each set of parameters.

\subsection*{Analytical Model of ``Implicit'' Density Dependence}
To develop a simple mean-field analytical model to describe the effect of the underlying viral dynamics on the phage diffusion, we imagine phage diffusing across a lawn of ``sticky'' penetrable disks. These disks are used to represent host bacteria cells that are able to adsorb phage for a period equivalent to the lysis time, after which the phage desorb and continue to diffuse. The disks do not pose a hindrance to the phage through steric interactions.

In this set-up, phage diffuse through a series of discrete steps, where phage move a certain distance with each step. In any given step, the probability that a phage will become adsorbed to one of the host bacteria is $p_{\alpha}\phi$, where $p_{\alpha}$ represents the probability of adsorbing when a phage encounters a host, and $\phi$ represents the fraction of all space occupied by the host bacteria. This is analogous to a Poisson point process, where events (adsorption) occur continuously and independently.  Therefore, the number of steps that a phage takes before becoming adsorbed to a host is exponentially distributed with mean $t_{ads}=\frac{1}{p_{\alpha}\phi}$. Consequently, over the period of time $T=t_{ads}+\tau_s$, where $\tau_s$ is the lysis time (in steps), the phage will only have on average actually moved for $t_{ads}$ of that time.

For long times (over many adsorption/desorption events), this process can be thought of as a hindered diffusion process with relative diffusion coefficient equal to
\begin{equation}
    \frac{D}{D_0}=\hat{D}_{imp}=\frac{t_{ads}}{t_{ads}+\tau_s}=\frac{1}{1+p_{\alpha}\tau_s\phi},
\end{equation}
where $D_0$ is the free diffusion coefficient, and $\hat{D}_{imp}$ is the relative density-dependent diffusion coefficient resulting from the hindrance posed by the underlying viral dynamics, which we have termed the ``implicit'' density-dependence.

This can be re-written in terms of the parameters used in the main text as
\begin{equation}
    \frac{D}{D_0}=\hat{D}_{imp}=\frac{1}{1+p_{\alpha}\tau_s\phi}=\frac{1}{1+AbK_{max}},
\end{equation}
where $A$ is a scaling parameter given by:
\begin{equation}
    A=\frac{p_{\alpha}\tau_s \phi}{\alpha \tau B_{max} b}.
\end{equation}
To compare the parameters used in the two descriptions, we can consider that in our mean-field model $\phi=b$. We can then use the fact that the term $\alpha B_{max}$ determines the rate at which phage are adsorbed when in contact with bacteria (as all space is filled with bacteria at $B_{max}$), and so, like $p_{\alpha} \tau_s$, the term $\alpha B_{max} \tau$ measures the total probability that phage will be adsorbed over a lysis time, assuming that the phage are always in contact with bacteria (i.e. $p_{\alpha} \tau_s = \alpha B_{max} \tau$). This leads to a value for $A$ of:
\begin{equation}
    A=\frac{p_{\alpha}\tau_s \phi}{\alpha \tau B_{max} b}=\frac{p_{\alpha}\tau_s}{\alpha \tau B_{max}} = 1.
\end{equation}
Consequently, the implicit density dependence can be written equivalently in terms of either parameters as:
\begin{equation}\label{eqn:implicit_diffusion_methods}
    \frac{D}{D_0}=\hat{D}_{imp}=\frac{1}{1+p_{\alpha}\tau_s\phi}=\frac{1}{1+bK_{max}}.
\end{equation}

\subsection*{Analytical Model Predicts Multiplicative Effects of Steric Interactions and Infection Dynamic}
The model introduced above can be modified to account for the presence of steric effects. In the absence of adsorption i.e., if excluded-volume interactions were the only hindrance to diffusion, the average fraction of steps successfully ``jumped'' by phage compared to the total attempted would be $1-\phi$, and we can define a relative diffusion coefficient $\hat{D}_{exp}=1-\phi$. Although this is an approximation as it does not take into account the fact that jumps may be correlated, which is why it deviates from the more precise Fricke's equation, it helps extending the analytical model in the previous section.

If we now introduce adsorption, as explained in the previous section, the average number of steps taken by phage before adsorbing to an obstacle will be $t_{ads}=\frac{1}{p_{\alpha}\phi}$. In the presence of steric interactions, only a fraction $1-\phi$ of these steps will be successful, so that $t_{succ}=\frac{1-\phi}{p_{\alpha}\phi}$. Thus, on average, the success rate of jumping if both adsorption and steric interactions are taken into account will be:
\begin{equation}
    \frac{D}{D_0}=\hat{D}_{exp+imp}=\frac{t_{succ}}{t_{ads}+\tau_s}=\frac{1-\phi}{1+p_{\alpha}\tau_s\phi},
\end{equation}
which is equivalent to the product $\hat{D}_{exp}\hat{D}_{imp}$, indicating that when both implicit and explicit effects are present, the total behaviour can be expressed as the product of both effects individually. Because in our model the functional form is explicitly input into the PDE system, the same argument holds if we replace the simplified $\hat{D}_{exp}=1-\phi$ with the more precise Fricke's law parameterised by our experiments, resulting in Eq.~\ref{eqn:VDM-}.

\subsection*{Implicit `Boost' to Diffusion}
This section will derive how the implicit diffusion coefficient in the UDM- can be thought of as a boost to the implicit diffusion coefficient in the UDM+. Here, let the implicit slow-down to diffusion in the UDM+ and UDM- be denoted by $\hat{D}_{imp+}$ and $\hat{D}_{imp-}$, respectively. Following our previous derivations, these rates are given by:
\begin{equation}
    \hat{D}_{imp_-^+}=\frac{1}{1+b_{_-^+}K_{max}},
\end{equation}
where $b_+=\frac{B+I}{B_{max}}$ and $b_-=\frac{B}{B_{max}}$, owing to the fact that infected cells are either adsorbing or non-adsorbing in the two models.

So as to compare to the dimensionless set of parameters used in the model (where $\overline{D}=1$ at the expansion front), we re-scale the implicit coefficients as:
\begin{equation}
    \overline{D}_{imp_-^+}=\frac{\hat{D}_{imp_-^+}}{\hat{D}_{imp_-^+}(f)}=\frac{1+fK_{max}}{1+\rho_{_-^+}fK_{max}},
\end{equation}
where $\rho_+=\overline{B}+\overline{I}$ and $\rho_-=\overline{B}$.

If we then compare the ratio of the two coefficients, it can be seen that
\begin{align}
    \frac{\overline{D}_{imp-}}{\overline{D}_{imp+}}&\approx\frac{1+fK_{max}}{1+\overline{B}fK_{max}}\frac{1+(\overline{B}+\overline{I})fK_{max}}{1+fK_{max}}\\
    &=\frac{1+(\overline{B}+\overline{I})fK_{max}}{1+\overline{B}fK_{max}}\\
    &=1+\frac{\overline{I}fK_{max}}{1+\overline{B}fK_{max}}.
\end{align}
The approximation arises from the assumption that the bacterial curves $\overline{B}$ and $\overline{I}$ are the same in both expansions. This is supported by the observation that the density profiles for $\overline{B}$ and $\overline{I}$ are very similar when compared across models (Fig.~\ref{fig:implicit_diffusion}b). Therefore we can see that we can write $\overline{D}_{imp-}$ in terms of $\overline{D}_{imp+}$ as
\begin{equation}
    \overline{D}_{imp-}\approx(1+\psi)\overline{D}_{imp+} \: ; \: \psi=\frac{\overline{I}fK_{max}}{1+\overline{B}fK_{max}}.
\end{equation}

\subsection*{Diffusion Profiles}
Fig.~\ref{fig:implicit_diffusion}b shows the proxy diffusion coefficients of each of the model variants as a function of position across the expansion front. To generate this, the population profiles $\overline{V}$, $\overline{B}$ and $\overline{I}$ of each of the model variants were taken at the final time step of the numerical solution (so as to be as close to the steady state as possible), and then aligned so that the half max points of the density profiles of uninfected bacteria ($\overline{B}=0.5$) coincide. These population curves were then used to determine the proxy dimensionless diffusion coefficients:
\begin{subequations}
\begin{equation}
    \overline{D}_{UDM+}=\overline{D}_{imp+},
\end{equation}
\begin{equation}
    \overline{D}_{UDM-}=(1+\psi)\overline{D}_{imp+},
\end{equation}
\begin{equation}
    \overline{D}_{VDM+}=\overline{D}_{exp}\overline{D}_{imp+},
\end{equation}
\begin{equation}
    \overline{D}_{VDM-}=(1+\psi)\overline{D}_{exp}\overline{D}_{imp+}.
\end{equation}
\end{subequations}

\subsection*{Stochastic Simulations}
Our simulation algorithm is carried out on a one-dimensional lattice (Fig.~\ref{fig:het_data}a). A finite number of lattice sites (demes), denoted by $i$, are distributed along a line, with each containing a fixed number of bacteria $B_i=B_0$. Each deme is also initialized with $V_i=100$ phage. In each time step, there is: a migration step in which a proportion of phage from each deme $i$, binomially sampled with $V_i$ trials and probability $m/2$ ($m=0.25$), are exchanged with each of its neighbors; an adsorption step in which the number of adsorbing phage is sampled in each deme from a binomial distribution with $B_i V_i$ number of trials, with success probability $\alpha$; and a lysis step in which each infected bacteria's state is advanced by one, and bacteria with state $\tau$ are labeled as lysed, and $\beta$ new phage are inserted into the deme.
The simulation box is periodically shifted with uninfected bacteria placed ahead of the population and demes with a steady state number of phage omitted and recorded. In this way the simulation box stays in the co-moving frame of the population. 

When the traveling wave is established, verified by convergence of the expansion velocity, all of the free and adsorbed phage are randomly labeled with one of two neutrals labels. The proportion of the population with each allele selected during the migration and adsorption step is found by binomial sampling with probability equal to the current allele fractions and the total number of events as described above. Upon lysis, all of the new phage released are labelled with the same marker as the phage which infected that bacteria.

\subsection*{Decay in Heterozygosity}
The average heterozygosity $H$ in the simulation box is given by
\begin{equation}
H = \frac{1}{M}\sum^M_i 2f_i(1-f_i),
\end{equation}
where $M$ is the total number of demes in the simulation box, and the fraction of the two alleles in deme $i$ are $f_i$ and $1-f_i$. Timesteps in our simulation are converted to generation time $T$ by noting that $T = 1/ \alpha B_0 + \tau$ timesteps is equal to the average time for an individual virus to be adsorbed and the infected bacterial cell to lyse.

It is expected that heterozygosity decays due to genetic drift in our simulations~\cite{Birzu2018,Birzu2019}. We expect that heterozygosity the $H(t)$, within a certain range of $t \in {(t_s,t_f)}$, will  approximately satisfy the relation $H(t) = Ae^{-\Lambda(t+B)} +C$, where $A$, $B$, and $C$ are constants. With variable transient periods, $A$ and $B$ are unknown, but we assume $C$ to be 0 (the heterozygosity will always decay to 0 as one of the alleles fixes). To estimate $\Lambda$, we simply take the natural log of our data, which we expect to be approximated by $\ln H(t) = \ln{A} -\Lambda(t+B)$. Combining constant terms, we can find $\Lambda$ by simply performing a linear fit to $\ln H(t)$:  
\begin{equation}
\ln H(t)= -\Lambda t + \mathrm{const}. 
\end{equation}
We can alternatively express $\Lambda$ in terms of an effective population size, $N_e$, where $N_e \equiv 1/\Lambda$~\cite{Hallatschek2008}. Following ~\cite{Birzu2018,Birzu2019}, the fit was performed for the average of 1000 simulations, with $t_i$ chosen such that $H(t_i)$ was as close to 0.1 as possible, and $t_f$ was chosen such that at least 5\% of simulations had non-zero values of $H(t_f)$ (Fig.~\ref{fig:het_data}b). Calculated $N_e$ was normalized by the measured average steady state viral population per deme $V_{ss}$ in the bulk of the established traveling waves to account for variable carrying capacity.

\subsection*{Time- and Length-Scales in Stochastic Simulations}
To express the $\tau$ and $\alpha B_0$ in our stochastic simulations in terms of real time units, we defined a spatial scale $L$ in our model such that $L^2= B^s_{0}/B^r_{0}=115 \mathrm{\frac{\mu m^2}{deme}}$, where $B^s_{0}$ and $B^r_{0}$  is the initial bacterial density in the simulations and in our experimental data, respectively. With the spatial scale fixed, we can find the equivalence between simulation timesteps and minutes $T$ using the viral diffusion constant in our simulations and our experimental data, $D^s$ and $D^r$ respectively. We first note that $D^s =m/2$ which is $0.125$ in all our simulations, and $D^r $ is function of $B^r_0 =B^s_{0}/L^2 $, in accordance to our fitted values in (Fig.~\ref{fig:Diffusion_Experiment}c).  We then have that
\begin{equation}
    T=\frac{D^s L^2}{D^r} = \frac{m L^2}{120 D^r(B^s_{0}/L^2)} \frac{\mathrm{min}}{\mathrm{timestep}},
\end{equation}
where $D^r$ is in units of $\frac{\mu m^2}{s}$. Given a value of $B^s_{0}$, we can use this equivalence to convert $\alpha B_0$ and $\tau$ to minutes.

\begin{acknowledgements}
We acknowledge help from R. Majed in performing bacterial transformations. DF thanks Kirill Korolev and Oskar Hallatschek for helpful discussions. MH acknowledges studentship funding from EPSRC under grant number EP/R513180/1. NK acknowledges funding from the Gates Cambridge Scholarship. TL acknowledges college grant from St Edmund Hall. This work was performed using resources provided by the Cambridge Service for Data Driven Discovery (CSD3) operated by the University of Cambridge Research Computing Service, provided by Dell EMC and Intel using Tier-2 funding from the Engineering and Physical Sciences Research Council (capital grant EP/P020259/1), and DiRAC funding from the Science and Technology Facilities Council.
\end{acknowledgements}

\appendix
\section{Phage Remain On Agar Surface}\label{app:agar_diffusion}
To verify that phage are not diffusing out of the approximately 2D plane and into the agar during the course of the diffusion experiments, additional tests were carried out. If significant diffusion into the agar was occurring, we would expect the number of phage at the surface to be reduced over time. As with the measurements of diffusion coefficient (see Methods), 35 ml omni-plates of 20 g/l (2\%) agar, with LB and 15 $\mu$g/ml chloramphenicol were prepared. Similarly, overnight liquid cultures of susceptible host (\textit{E. coli} eMTH43) were grown from single colonies at 37 \degree C in LB with  15 $\mu$g/ml chloramphenicol.

Then, 1.5 ml of stock bacteriophage T7 diluted in LB was spread across the plate with glass beads, such that the plate should contain a countable (order of tens) number of phage.

\begin{figure}[h]
\includegraphics[width=0.45\textwidth]{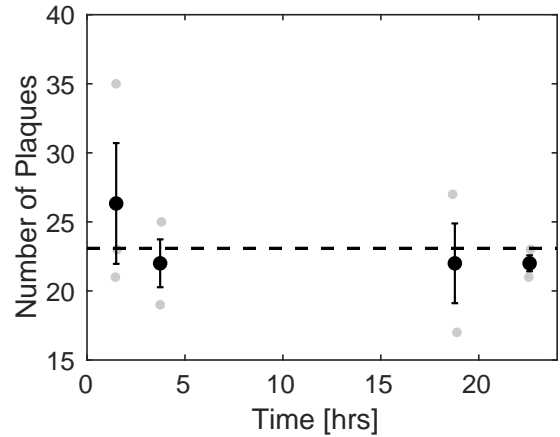}
\caption{\label{fig:agar} Plaques counted on the surface of a 2\% agar plate as a function of time after the phage were originally added to the plate. Data for the four time points measured (black) and their corresponding errors are calculated from three replicates (grey). The dashed line indicates the mean number of plaques counted across time-points. It can be seen that the data is consistent with a constant number of phage being recovered over time, indicating that phage diffusing into the agar is very limited.}
\end{figure}

After a set time period (approximately 1.5, 4, 19 or 23 hours), 100 $\mu$l of overnight eMTH43 was mixed with 10 ml of molten 7 g/l (0.7\%) agar and poured over the surface of the plate and left on a lab bench overnight. Any phage that were located on the surface of the 2\% agar at the time in which the 0.7\% agar and susceptible eMTH43 was poured on top is expected to be able to infect the susceptible host and result in a plaque. By counting the number of plaques visible in the 0.7\% agar the following morning, we are able to determine whether a signficant amount of phage diffuses into the 2\% agar plate over a 24 hour period as these phage would not be able to form a plaque.

The results from this test (Fig.~\ref{fig:agar}) clearly show that there is no significant reduction in the number of phage recovered from the surface of the plate over a roughly 24 hour period (the period over which diffusion measurements were gathered). We believe that this is because the pore size of our 2\% agar is small enough to significantly limit diffusion. Indeed, similar 2\% agarose substrates have been shown to have a pore size of under 80 nm \cite{Narayanan2006}, which is comparable with the size of T7.

\section{Monte Carlo Simulations of Diffusion}\label{app:MonteCarlo}
Monte Carlo simulations of point-like phage diffusing across a lawn of penetrable ``sticky'' disks, representing adsorbing host bacteria in the absence of steric effects were performed. The simulations are carried out in a 2D box with periodic boundary conditions, populated with circular obstacles of fixed diameter $d_{obs}$, representing a lawn of host bacteria cells. Obstacles are generated with random center co-ordinates within the boundary of the lawn until a fraction $\phi$ of the area of the lawn is occupied (obstacles may overlap). We use \textit{sticky penetrable} obstacles to mimic the implicit hindrance to diffusion due to infection dynamics. A hundred of point-like tracers (representing phage) are placed on the lawn and diffuse in discrete steps. At each step, every free tracer proposes a new random co-ordinate within a circular region of radius $r_{step}$ of its current position. If there is no obstacle at the new co-ordinate, the tracer jumps to the new position. If, however, there is an obstacle at the new co-ordinate, the tracer jumps to the new position and, with probability $p_{\alpha}$, adsorbs to that obstacle. Upon successful adsorption, the tracer remains bound to the obstacle for a number of steps $\tau_{s}$ (representing the viral incubation period). 

\begin{figure}[h]
\includegraphics[width=0.45\textwidth]{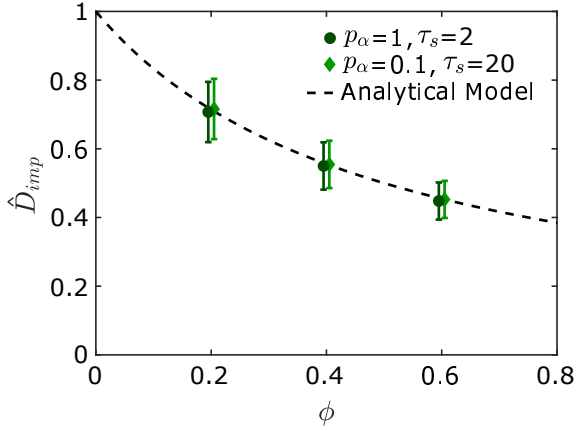}
\caption{\label{fig:MonteCarlo_sims} Relative effective diffusion coefficient obtained from Monte Carlo simulations of point-like phage diffusing through a field of penetrable ``sticky'' obstacles. These obstacles result in a reduction in the effective diffusion coefficient of the phage in agreement with the mean-field analytical model of implicit density-dependence Eq.~\ref{eqn:implicit_diffusion_methods} (see Methods). Simulation data offset slightly for clarity.}
\end{figure}

The size of the lawn is $1000\,d_{obs}$, and $r_{step}=5\,d_{obs}$. We use 100 phages for each simulation, and run 100 simulations. The simulations are run for 3 million steps, to observe equilibration of the diffusive behavior. After a brief initial transient period, we plot the mean square displacement averaged over all tracer particles in each simulation as a function of time and fit a linear function. The slope of this function is equivalent to $4D$ where $D$ is the diffusion coefficient. To randomise the position of the host, we run 100 independent simulations. The final diffusion coefficient is estimated by calculating the average and the standard deviation of the diffusion coefficients across these simulations. The relative diffusion coefficient $\hat{D}_{imp}$ is then determined by dividing this results with the theoretical free diffusion coefficient given by the simulation parameters.

These results show that the hindrance posed by the underlying viral dynamic is equivalent to an ``implicit'' density-dependent reduction in diffusion $\hat{D}_{imp}$ (Fig.~\ref{fig:MonteCarlo_sims}). We find quantitative agreement with the mean-field analytical model discussed in the main text (Eq.~\ref{eqn:implicit_diffusion} and Methods), both in terms of the dependence on $\phi$, and that the behaviour does not depend on $p_{\alpha}$ and $\tau_s$ independently, but rather depends only on the product $p_{\alpha}\tau_s$. This is in agreement with the fact that the behaviour in the UDMs depends only on $K=\alpha \tau B_0$.

\section{Ratio of Explicit to Implicit Effects}\label{app:effects_ratio}
\begin{figure}[h]
\includegraphics[width=0.45\textwidth]{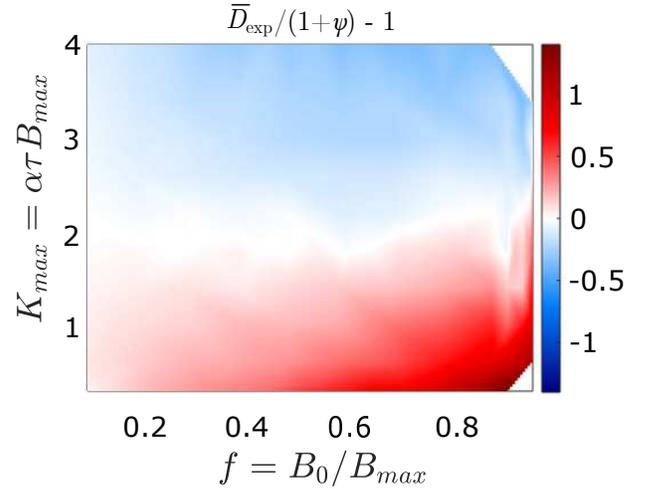}
\caption{\label{fig:effects_ratio} The ratio of the explicit boost to diffusion $\overline{D}_{exp}$ to the implicit boost to diffusion $(1+\psi)$, as a function of $f$ and $K_{max}$.}
\end{figure}

Here we present a plot of the ratio of the explicit boost to diffusion $\overline{D}_{exp}$, when transitioning from UDM to VDM, to the implicit boost to diffusion $(1+\psi)$ when transitioning from UDM+ to UDM-, or from VDM+ to VDM-, as a function of $f$ and $K_{max}$. This is obtained by determining the strength of each effect at the front position where the phage population is 3/4 times the steady state population $\overline{V}_{ss}$. It can be seen in Fig.~\ref{fig:effects_ratio} that the implicit boost is dominant at large $K_{max}$, while the explicit boost dominates at low $K_{max}$.

\section{The Impact of Burst Size} \label{app:burst}
\begin{figure}[h]
\includegraphics[width=0.45\textwidth]{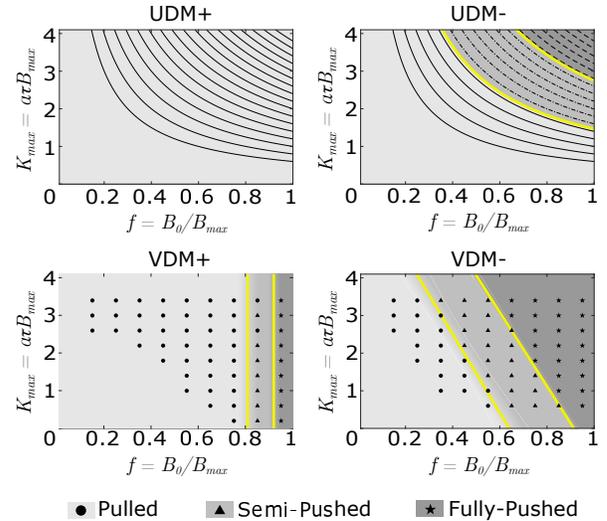}
\caption{\label{fig:phase_Y20} Phase diagrams showing the expansion types for the four model variants as a function of $f$ and $K_{max}$, $\beta$=20 throughout. As can be seen by comparison to Fig.~\ref{fig:phase}, the qualitative behaviour of the transitions in each of the models remains the same, and can be characterised using the same variables: $K_{s,p}$ for the UDM-; $f_{s,p}$ for the VDM+; $m_{s,p}$ and $a_{s,p}$ for the VDM-. As before, lines in the UDMs, and data points in the VDMs indicate the parameter combinations for which numerical integration was performed, and velocities calculated. Transition boundaries (yellow lines) are inferred from the data points calculated.}
\end{figure}

\begin{figure*}
\includegraphics[width=0.99\textwidth]{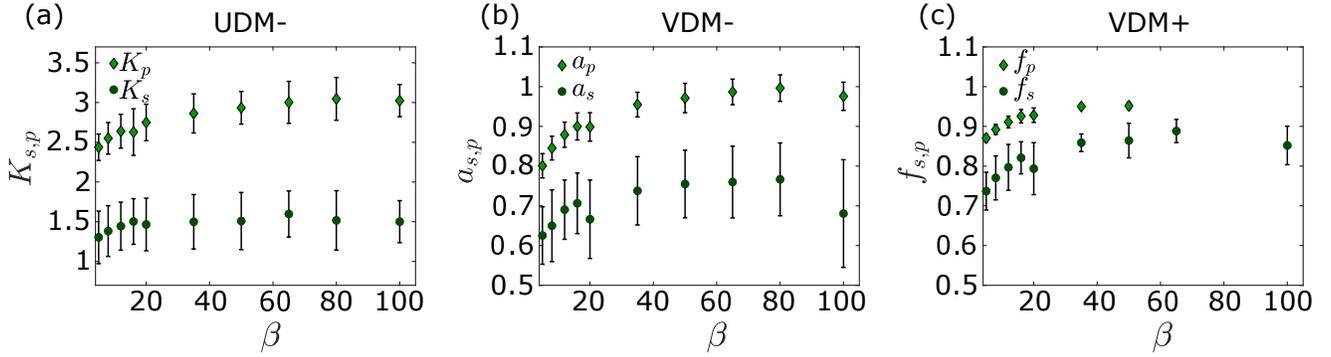}
\caption{\label{fig:burst} Behavior of the critical parameters describing the location of the transitions (Fig.~\ref{fig:phase} and Fig.~\ref{fig:phase_Y20}) as a function of burst size $\beta$. (a): Critical values $K_s$ and $K_p$ in the UDM-. (b): Critical values $a_s$ and $a_p$ in the VDM-, calculated from transition locations at $K_{max}$=2.2, assuming the gradients of the transitions $m_s$ and $m_p$ are approximately constant when varying $\beta$, to maintain computational feasibility. (c): Critical values $f_s$ and $f_p$ in the VDM+, similarly calculated from transition locations at $K_{max}$=2.2. For $\beta>50$ we did not observe transitions to fully pushed waves in the parameter regime of $f\leq0.95$ investigated. Extending the parameter regime was computationally unfeasible. In each case, to determine the error bars, we assume a 1\% error in the model velocities as before, and use this to determine the resultant shift in the transition parameters.}
\end{figure*}

\begin{figure}[h]
\includegraphics[width=0.45\textwidth]{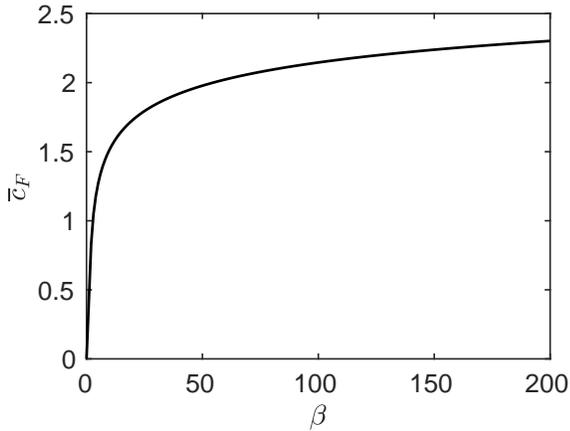}
\caption{\label{fig:CvsY} Spreading velocity of the linearised model for $K=1.0$ as a function of burst size $\beta$ (in the linearised model, $K$ and $\beta$ are the only independent parameters).}
\end{figure}

In Fig.~\ref{fig:phase_Y20} we present phase diagrams such as in Fig.~\ref{fig:phase}, but for a burst size $\beta$=20 instead of $\beta$=50. It can be seen by comparing both Figures that the qualitative behaviour of the transitions remains unchanged: in the UDM-, transitions are characterised by constant values of $K_s$ and $K_p$; in the VDM-, transitions are approximately straight lines characterised by gradients $m_s$ and $m_p$, and intercepts $a_s$ and $a_p$ ($f=m_{s,p}K_{max}+a_{s,p}$); in the VDM+, transitions are largely independent of $K_{max}$, and occur at critical values $f_s$ and $f_p$.

As it seems the behaviour can be characterised in the same manner when burst size is changed, we simplify our examination by focusing only on how the burst size alters these specific characteristic parameters. Rather than attempting to produce the whole phase diagram for each of the models at various $\beta$, as this is very computationally intensive when $\beta$ is either small or large, we instead choose a specific $K_{max}$ value, and for various values of $\beta$, we vary $f$ at this $K_{max}$ value. From this, the parameters $K_{s,p}$ and $f_{s,p}$ as a function of $\beta$ can be easily obtained (i.e. the parameters describing the UDM- and VDM+ transitions respectively). 

To simplify our investigation of the behaviour in the VDM-, and limit the number of computationally intensive calculations required, we assume that $m_{s,p}$ are constant with burst size, and using data from one specific $K_{max}$ value we calculate how $a_{s,p}$ vary as a result. In the two cases where the full phase diagrams were computed, the transition gradients were calculated as $m_s=-0.125(5), -0.091(11)$ and $m_p=-0.097(5), -0.101(2)$ for $\beta=50, 20$, indicating agreement to within $2\sigma$ and $1\sigma$ respectively.

We find that while the general shape of the transitions for the model variants does not depend on $\beta$ (Fig.~\ref{fig:phase_Y20}), the exact location of the transitions are affected (Fig.~\ref{fig:burst}). The dependence of all of the transition parameters on $\beta$ is similar across models. Above $\beta \approx$ 40, the parameters exhibit only a weak dependence on burst size, whereas when $\beta$ decreases below this value, the transition parameters also decrease, increasing the parameter range of $K_{max}$ and $f$ in which we observe a pushed wave.

This behaviour qualitatively matches the dependence of the spreading velocity of the linearised model $\overline{c}_F$ on burst size (Fig.~\ref{fig:CvsY}). While $f$ and $K_{max}$ determine the ability of phage in the bulk to catch up to the front and contribute to the dynamics, either due to explicit or implicit hindrance to diffusion, $\beta$ only contributes to the phage growth rate and, as a result, the velocity of the front. At lower values of $\beta$, the spreading velocity is greatly reduced as the limited number of phage released at the tip after each lysis event struggle to clear the host cells around them. This allows the phage in the back to catch up more easily, regardless of the mechanism, and contribute to the expansion. As burst size is increased however, the opposite is true, although the velocity gains that come with increased $\beta$ become increasingly marginal, as the uninfected host within the vicinity of recently lysed cells become saturated with newly released phage.

\section{Bacterial Growth} \label{app:growth}
The model we present in the main text assumes that the host bacteria are not growing. While this certainly can be the case, it is by no means always true, and so it is natural to ask how our results are affected by a growing host population. To this end, we modify our reaction-diffusion model to include a logistic growth term:
\begin{subequations}\label{eqn:Model_Growth}
\begin{equation}
\pdv{B}{t} = -\alpha VB + \mathbf{r_0 B\left(1-\frac{B+I}{B_0}\right)},
\end{equation}
\begin{equation}
\pdv{I}{t} = \alpha VB - \alpha V_{t-\tau}B_{t-\tau},
\end{equation}
\begin{equation} 
\pdv{V}{t} = \pdv{}{x} \left(D \pdv{V}{x} \right) - \alpha VB - \alpha^* VI + \beta\alpha V_{t-\tau}B_{t-\tau},
\end{equation}
\end{subequations}
where terms in bold indicate new terms, and $r_0$ is the growth rate of the bacteria at low densities. We estimate this growth rate by assuming a doubling time of bacteria equal to 30 mins (typical of T7 host \textit{E. coli}), resulting in $r_0=$ ln(2)/30 min$^{-1}$. Note, that in each case the carrying capacity is equal to the initial bacterial density $B_0$, meaning that growth only occurs across the front region. This represents the situation that would occur in natural environments where bacteria far from the front have grown to a stationary density (set by $B_0$) by the time the front arrives. Additionally, we assume that the replication of infected hosts is negligible due to the burden caused by producing new phage.

\begin{figure}[h]
\includegraphics[width=0.45\textwidth]{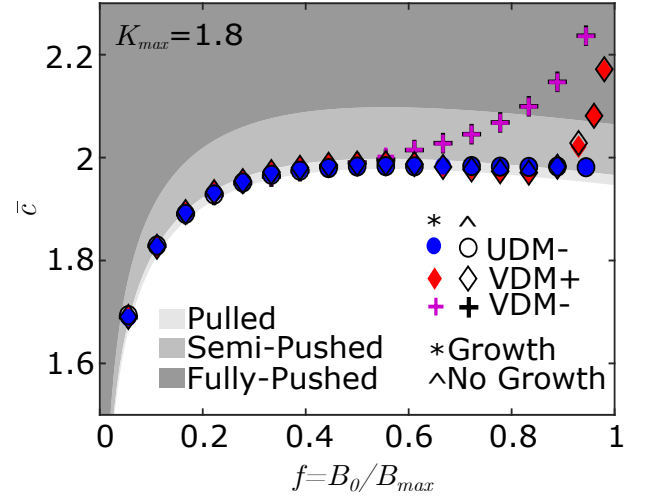}
\caption{\label{fig:growth} Dimensionless front velocity $\overline{c}$ as a function of bacteria fraction $f$, with shaded regions indicating the different expansion types, when bacteria grow logistically as described in Eqs.~\ref{eqn:Model_Growth}. As in Fig.~ \ref{fig:fronts_transitions}c, parameters are chosen to represent typical T7 expansions with $\beta=50$, $\tau$=18 mins, and $\alpha B_{max}$=0.1 min$^{-1}$~\cite{Yin1992, Jones2012, Rioja2015}, corresponding to $K_{max}=1.8$ in our model. Additionally, a doubling time of 30 mins has been assumed to calculate the growth rate of the host bacteria. By comparison to the results obtained in the absence of bacterial growth (as in Fig.~ \ref{fig:fronts_transitions}c), it can be seen that bacterial growth has no effect on the transition to pushed expansions.}
\end{figure}

It can be seen in Fig.~\ref{fig:growth} that the introduction of this bacterial growth term has no discernible effect on the type of expansion that occurs in the parameter space describing typical T7 expansion through host \textit{E. coli}.

\section{Cell Debris} \label{app:debris}
Another aspect of the system which our model thus far neglects is the possibility that some of the debris from lysed cells could trigger phage adsorption. If this were the case, it would likely result in a reduction of the parameter space corresponding to pushed waves, as there would be fewer phage in the bulk that were able to catch up to the front. To explore this prospect further, we modify our reaction-diffusion model to account for this possibility:
\begin{subequations}\label{eqn:Model_Debris}
\begin{equation}
\pdv{B}{t} = -\alpha VB,
\end{equation}
\begin{equation}
\pdv{I}{t} = \alpha VB - \alpha V_{t-\tau}B_{t-\tau},
\end{equation}
\begin{multline}
\pdv{V}{t} = \pdv{}{x} \left(D \pdv{V}{x} \right) - \alpha V(B+I) + \beta\alpha V_{t-\tau}B_{t-\tau} \\ - \mathbf{\alpha V(B_0-(B+I))d_f},  
\end{multline} 
\end{subequations}
where terms in bold indicate new terms, and $d_f$ controls the fraction of lysed cells that are capable of adsorbing phage (or alternatively, how able the cell debris is to adsorb phage in comparison to the cells which generated it). Note that we have limited the model to the case where adsorption to infected hosts occurs, as it seems unlikely that phage would be unable to adsorb to infected hosts but would be able to adsorb to the debris from those hosts (VDM+).

\begin{figure}[h]
\includegraphics[width=0.45\textwidth]{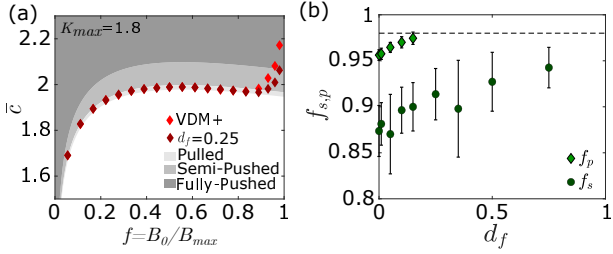}
\caption{\label{fig:debris} (a): Dimensionless front velocity $\overline{c}$ of the VDM+ as a function of bacteria fraction $f$ and debris fraction $d_f$, with shaded regions indicating the different expansion types. As in Fig.~ \ref{fig:fronts_transitions}c, parameters are chosen to represent typical T7 expansions with $\beta=50$, $\tau$=18 mins, and $\alpha B_{max}$=0.1 min$^{-1}$~\cite{Yin1992, Jones2012, Rioja2015}, corresponding to $K_{max}=1.8$ in our model. (b): Dependence of the critical bacteria fractions $f_s$ and $f_p$, where transitions occur, on $d_f$. The dashed line indicates the upper bound of $f$ for which results were computed. We expect fully pushed transitions to continue to occur above this threshold for some time as $d_f$ is increased.}
\end{figure}

It can be seen in Fig.~\ref{fig:debris}a that while the introduction of adsorbing cell debris does not qualitatively change the behaviour of the model (i.e. that transitions to pushed waves occur at high bacterial fractions in the VDM+), it does have a slight impact on the location of these transitions (Fig.~\ref{fig:debris}b). As the fraction of cell debris capable of adsorbing phage $d_f$ increases, the critical bacteria fractions $f_s$ and $f_p$ where transitions occur also increase. This result confirms our intuition: a larger $d_f$ will cause a greater reduction in free phage in the bulk of the plaque, thereby reducing the number of phage available to catch up to the front and contribute to the expansion dynamics.

\begin{figure}[h]
\includegraphics[width=0.45\textwidth]{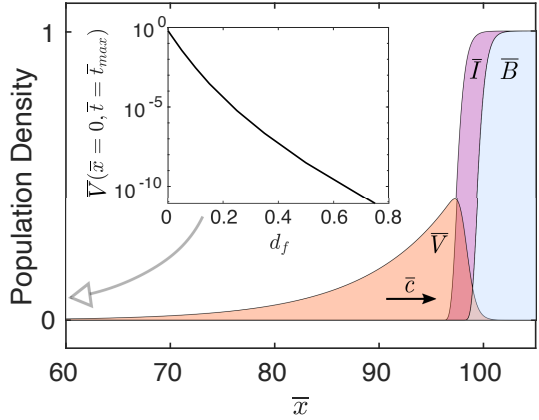}
\caption{\label{fig:debris2} Population concentrations $\overline{B}$, $\overline{I}$ and $\overline{V}$ at time $\overline{t}=\overline{t}_{max}$ for a debris fraction $d_f$=0.25. The front is propagating with dimensionless velocity $\overline{c}$ to the right. It can be seen that behind the expansion front the density of the viral population is quite low. Inset shows the density of the viral population at $\overline{x}$=0 and $\overline{t}=\overline{t}_{max}$ as a function of $d_f$ (i.e. it indicates the viral density in the centre of the plaque).}
\end{figure}

While increasing the rate at which cell debris adsorbs phage does result in `less pushed' expansions, we can see in Fig.~\ref{fig:debris2} that the resulting depletion of free phage in the bulk of the expansion is even more significant. Indeed, after an expansion of 50 lysis times (15 hours if $\tau$=18 mins) we find that there are approximately 10$^5$ fewer phage in the bulk than would be found at the expansion front when $d_f$=0.25, and approximately 10$^{10}$ when $d_f$=0.6. These values are clearly unrealistic for T7, as in the laboratory phage can be easily recovered from the centre of a plaque by simply stabbing with a needle. Therefore, if we assume that the steady state phage population at the front is approximately $\beta B_0$=100 $\mu$m$^{-2}$ ($\beta$=100, $B_0$=1 $\mu$m$^{-2}$), and then \textit{very} conservatively assume that easy recovery would require at least a phage density of 10$^3$ mm$^{-2}$, then we can conclude that any $d_f$ leading to a central density reduction of more than 10$^{5}$ is physically unrealistic (i.e. $d_f\geq0.25$).

Given that transitions to both semi-pushed and fully-pushed waves occur in our model up to a debris fraction $d_f$=0.25 (which we believe to be the \textit{very} upper bound of physically realistic behaviour) with only a minor shift in the bacterial fractions where these transitions occur, we can conclude that this effect, should it occur, has only a minor impact on our results.

\section{Heterozygosity Decay Without Normalization}\label{app:Het_extra}
\begin{figure}[h!]
\includegraphics[width=0.45\textwidth]{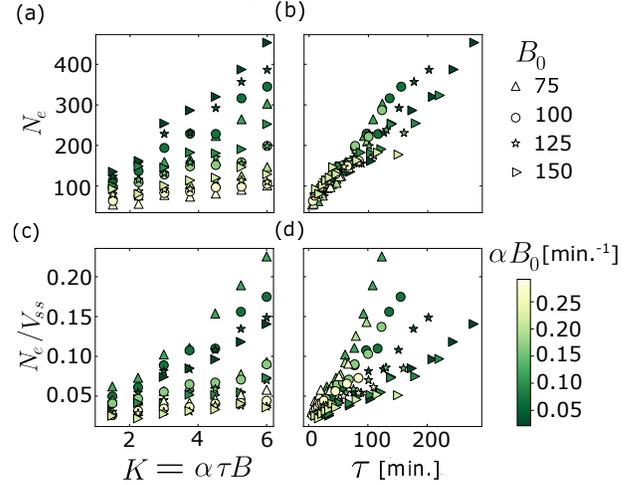}
\caption{\label{fig:het_supp} Effective population size $N_e$ over a range of $B_0$ values shown: (a) unnormalised and as a function of $K$; (b) unnormalised and as a function of $\tau$; (c) normalised by $V_{ss}$ and as a function of $K$; (d) normalized by $V_{ss}$ and as a function of $\tau$. Color indicates $\alpha B_0$ in min$^{-1}$. $B_0$ and $\alpha B_0$ were varried as they are the quantities one would measure independently in experiments. Effective population size is mainly controlled by $\tau$ as in Fig.~\ref{fig:het_data}d, over the range parameters examined. Errors due to linear fit of heterozygosity over time curve are negligibly small and not shown.}
\end{figure}

In addition to the results presented in the main text (Fig.~\ref{fig:het_data}), here we present the effective population size observed in our stochastic simulations for various values of $B_0$, both before and after normalisation by the steady-state population size $V_{ss}$ (Fig.~\ref{fig:het_supp}). This normalisation aimed to facilitate a comparison with previous theoretical studies where the carrying capacity of the population was kept constant. As was discussed in the main text however, the parameters in those studies (including carrying capacity) could be independently controlled, whereas in our system the steady state population size is an emergent property that depends both on the phage infection parameters, and the given aspects of the model variant in question. It should therefore be noted that the phage population size behind the expansion front will not always be directly comparable to the carrying capacity as used in these previous studies - take for example the case where phage can adsorb to cell debris (Appendix~\ref{app:debris}), resulting in the depletion of phage behind the front over time.

\section{Measuring Lysis Time on Solid Media}\label{app:solid_lysis}
\begin{figure}[h!]
\includegraphics[width=0.45\textwidth]{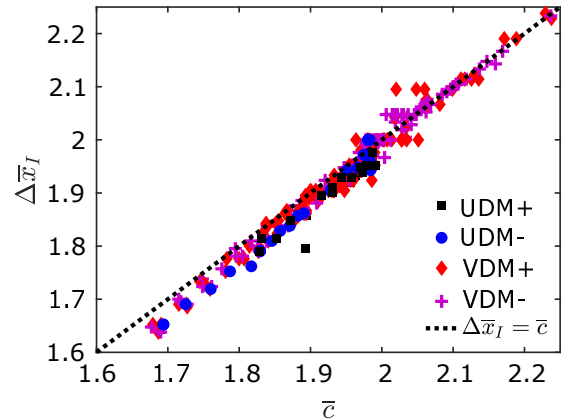}
\caption{\label{fig:Iw} Dimensionless width $\Delta\overline{x}_I$ shown to be equal to $\overline{c}$ for all models across the range of $f$ and $K_{max}$ investigated in Fig.~\ref{fig:phase}, corresponding to an expansion that travels the width of the infected region every lysis time. We attribute the small discrepancy observed at lower values to the limited convergence of the front profile to its steady-state because of trade-offs between precision and computational cost.}
\end{figure}

As discussed in the main text, our diffusion coefficient experiments raise the question of whether other life history parameters could also depend on the surrounding environment. With this in mind, our findings suggest a way to measure lysis time on solid media and over time during a plaque expansion. Lysis time is traditionally measured in liquid media, since it requires periodic and precise sampling of the phage population, which is challenging to perform on agar plates. The models presented in this paper, which assume a deterministic lysis time, produce a front which expands by the width of the infected region $\Delta\overline{x}_I$ during one lysis time interval. As a result, the dimensionless infection region width $\Delta\overline{x}_I$ is equivalent to the dimensionless velocity $\overline{c}$, as predicted in all the models and confirmed by the numerics (Fig.~\ref{fig:Iw}). By utilising phage engineered to result in fluorescent infected cells \cite{Vidakovic2018}, imaging a growing plaque of such phage over time in both fluorescent and bright-field channels should yield information about the distributions of infected and uninfected bacteria (see Fig.~\ref{fig:fronts_transitions}a), and enable determination of the dimensional equivalent of $\Delta\overline{x}_I$. By simultaneously measuring the velocity of the expansion, the lysis time in solid media and its variation during the course of an expansion could be estimated.

\newpage
\bibliography{references.bib}

\end{document}